\documentclass[12pt]{spieman}
\usepackage{amsmath,amssymb,amsbsy,amsfonts}
\usepackage{mathrsfs}
\usepackage{xcolor}
\usepackage{float}
\usepackage{graphicx}
\usepackage[algo2e,ruled,vlined]{algorithm2e}
\usepackage{wrapfig,lipsum,booktabs}
\usepackage{cite}
\usepackage{amsmath,amssymb,amsfonts}
\usepackage{graphicx}
\usepackage{textcomp}
\usepackage{mathtools}

\newcommand{\bs}[1]{\ensuremath{\boldsymbol{#1}}}

\newcommand{\R}{\mathbb{R}}

\newcommand{\rr}{\mathbf{r}}
\newcommand{\sos}{c}

\newcommand{\data}{ \bs d }

\DeclareMathOperator*{\argmin}{argmin}

\usepackage[math]{cellspace}
\usepackage{graphicx}
\usepackage{algorithm}
\usepackage{algpseudocode}

\usepackage{float}
\usepackage{enumerate}
\usepackage{url}
\usepackage{multirow}
\usepackage{wrapfig,lipsum,booktabs}
\usepackage{lineno}

\newcommand{\tumorlabelmap}{\boldsymbol{m}}

\title{Learned Correction Methods for Ultrasound Computed Tomography Imaging Using Simplified Physics Models}

\author[a,b,c]{Luke Lozenski}
\author[c]{Hanchen Wang}
\author[e]{Fu Li}
\author[e]{Mark A. Anastasio}
\author[d]{Brendt Wohlberg}
\author[f]{Youzuo~Lin}
\author[a,*]{Umberto Villa}

\affil[a]{The University of Texas at Austin, Oden Institute for Computational Engineering and Sciences, Austin, TX 78712 USA}
\affil[b]{Washington University in St. Louis, Department of Electrical and Systems Engineering, St$.\,$Louis, St$.\,$Louis, MO 63130 USA}
\affil[c]{Los Alamos National Laboratory, Applied Mathematics and Plasma Physics Group, Theoretical Division, Los Alamos National Laboratory, Los Alamos, NM  87545 USA}
\affil[d]{Los Alamos National Laboratory, Information Sciences Group,  Computer, Computational, and Statistical Sciences Division, Los Alamos, NM  87545 USA}
\affil[e]{University of Illinois Urbana-Champaign, Computational Imaging Science Laboratory, Department of Bioengineering, Urbana, IL, 61801 USA}
\affil[f]{University of North Carolina at Chapel Hill, School of Data Science and Society, Chapel Hill, NC 27599, USA}

\newcommand{\absdiv}[1]{
  \par\addvspace{.5\baselineskip}
  \noindent\textbf{#1}\quad\ignorespaces
}

\begin{document}

\maketitle
\vspace{-.11cm}
\begin{abstract}
\absdiv{Purpose:} Ultrasound computed tomography (USCT) is an emerging modality for breast imaging.  Image reconstruction methods that incorporate accurate wave physics produce high resolution quantitative images of acoustic properties but are computationally expensive.  The use of a simplified linear model in reconstruction reduces computational expense at the cost of reduced accuracy. This work aims to systematically compare different learning approaches for USCT reconstruction utilizing simplified linear models. 

\absdiv{Approach:}  This work considered various learning approaches to compensate for errors stemming from a linearized wave propagation model: correction in the data and image domains. The resulting image reconstruction methods are systematically assessed, alongside data-driven and model-based methods, in four virtual imaging studies utilizing anatomically realistic numerical phantoms. Image quality was assessed utilizing relative root mean square error (RRMSE), structural similarity index measure (SSIM), and a task-based assessment for tumor detection.

\absdiv{Results:} Correction in the measurement domain resulted in images with minor visual artifacts and highly accurate task performance. Correction in the image domain demonstrated a heavy bias on training data, resulting in hallucinations, but greater robustness to measurement noise. Combining both forms of correction performed best in terms of RRMSE and SSIM, at the cost of task performance. 

\absdiv{Conclusion:} This work systematically assessed learned reconstruction methods incorporating an approximated physical model for USCT imaging. Results demonstrated the importance of incorporating physics, compared to data-driven methods. Learning a correction in the data domain led to better task performance and robust out-of-distribution generalization compared to correction in the image domain. 
\end{abstract}


{\noindent \footnotesize\textbf{*}Umberto Villa,  \linkable{uvilla@oden.utexas.edu} }

\section{Introduction}
Ultrasound computed tomography (USCT) is an emerging medical imaging technology that can provide high-resolution estimates of tissue acoustic properties by utilizing ultrasound and tomographic principles\cite{natterer1995propagation, duric2005development}. USCT image formation relies on the interactions of acoustic wave signals with biological tissues.  Quantitative reconstructions of a tissue's acoustic properties can be achieved from USCT data via a variety of computational methods modeling these acoustic interactions\cite{li2009vivo, huthwaite2011high} and contain significant diagnostic value for breast cancer \cite{duric2007detection, schreiman1984ultrasound}. 

Full waveform inversion (FWI) \cite{wang2015waveform, virieux2009overview} is one such computational method that can be utilized for high-resolution estimation of a tissue's acoustic properties but requires solving a computationally expensive, nonconvex optimization problem. The computation expense of this inversion can be reduced utilizing a simplified mathematical model such as the Born approximation\cite{greenleaf1983computerized, lebras1988iterative}, which discards higher-order scattering effects, but sacrifices the primary benefit of accuracy.

This work assesses multiple machine learning methods to correct for model mismatch during FWI reconstruction utilizing the Born approximation. The methods considered include the application of a convolutional neural network (CNN) for artifact correction in the image domain after reconstruction utilizing an approximate physics model, the introduction  of a CNN for data correction in the measurement domain before reconstruction, and the combination of both of these forms of correction in a dual correction method. These image reconstruction methods utilizing an approximated physics model and machine learning are also compared to a purely data-driven reconstruction method that directly learns a mapping from measurement data to reconstructed images. These types of learned image reconstructions for FWI is an area of rapid development \cite{LinFengTheilerEtAl24}.

Utilizing an approximate physical model for reconstruction often results in artifacts in the image domain due to model mismatch. Previous works have demonstrated that machine and deep learning methods are well suited for image-to-image artifact correction and removal in a wide range of applications\cite{galteri2017deep, dong2015compression, kwon2015efficient, wang2022artifact, xu2022learning}. These approaches have been extended to a range of applications for artifact correction in biomedical imaging applications \cite{cui2016learning, khan2021variational, liu2021learning, lu2021artifact}, including for correcting artifacts due to mismatch in physical models related to measurement generation and image reconstruction \cite{jeong2023investigating, poimala2024compensating, gjesteby2017deep, lee2020review, abdoli2012metal}. This work assesses an image-to-image learned reconstruction method that leverages the Born approximation to construct the preliminary estimate of the object and a CNN for artifacts correction. 
These image-to-image methods learn an implicit image prior based on the set of training images, which can result in images that lead to realistic-looking reconstructed features but are heavily biased by the set of images used for training.

Previous approaches have also proposed deep learning methods that correct or enhance measurement data before image reconstruction\cite{yu2020deep, park2018ct, ghani2018deep, arabi2021assessment, sun2021coil}. Similarly, other approaches have examined methods for learned corrections in physical models to enable reconstruction utilizing a simplified forward model corrected by a neural network\cite{lunz2021learned}. This work assesses the use of a CNN as a means of learned measurement correction, or data correction. Specifically, the CNN maps USCT measurements --- either experimentally acquired or synthetically  generated by solving the wave equation --- to idealized measurements generated utilizing the Born approximation. Applying this correction exploits the convexity and computational efficiency of Born inversion, while reducing modeling errors. 

Correction on the measurements before inversion can be viewed as a form of pre-processing and avoids some of the inherent issues of image-to-image deep learning based image reconstruction methods that apply a learned correction in the image domain after inversion (post-processing). First, this measurement correction is based on well-posed mathematical relationships as opposed to artifact correction which is based on an image prior that is implicitly learned from the training examples; therefore data correction methods avoid inducing direct bias in the distribution of reconstructed images based on the learned image prior from training. Second, the number and size of measurements collected are often much larger than the size of reconstructed images and thus provide richer training sets for learned methods.   These two benefits mean that measurement correction is, in general, an easier map to learn and more generalizable compared to artifact correction in the image domain. 

Four simulation studies were performed to systematically compare these machine learning-based reconstruction methods for USCT imaging problems. The objects used in this study were a large set of anatomically realistic numerical breast phantoms with stochastically assigned acoustic properties within physiological ranges\cite{PhantomGen}. Ultrasound measurements were numerically simulated, assuming a stylized 2D ring-array USCT system with 64 sources,  256 receivers, and a 96 mm radius. The neural networks used for correction in the data and/or image domain utilized a U-net architecture\cite{ronneberger2015u}. 
A collection of 1435 examples was used to create disjoint training and testing sets for each of the numerical studies. Each example in the collection consisted of a speed of sound image  (214$\times$214 pixels), the corresponding USCT measurements simulated by use of the high-fidelity physics model (the wave equation), and the synthetic measurements generated via an approximate mathematical model (the Born approximation).

The results of these studies were systematically analyzed utilizing both physical and task-based metrics of image quality. Physical metrics of image quality reported were the mean square error to quantify overall image accuracy and structural similarity index measure to quantify perceptual image quality. Task-based assessment of image quality\cite{zhang21, adler22, he2013model} considered tumor detection and localization.
Specifically, in this work, a neural network based numerical observer was trained to perform a tumor detection and localization task. Several numerical observers are trained, one for each reconstruction method, using the same training set employed to train the image reconstruction method.  A receiver operator characteristic analysis was performed on reconstructed images from the testing set and the resulting area under the curve is reported as a figure of merit. The use of task-based metrics allows for  clinically relevant assessment of the considered learning methods as physical metrics of image quality do not directly correlate with the usefulness of a reconstructed image for performing a task of interest\cite{barrett93, christianson15, li21}, such as tumor detection and localization. 

The remainder of this paper is structured as follows. In Section~\ref{sec:background}, an imaging operator for USCT incorporating the wave equation or its Born approximation is introduced along with the formulation of model-based image reconstruction as an optimization problem. A brief overview of the basic principles of task-based assessment of image quality is also provided in this section. In Section~\ref{sec:method}, multiple methods utilizing machine learning for image reconstruction are presented. 
In Section \ref{sec:numerical_studies}, four numerical studies are designed to systematically compare the considered learned reconstruction methods with respect to the FWI method, which serves as a reference. In Section \ref{sec:results}, the results of these numerical studies are presented and discussed. Section \ref{sec:conclusion} presents the conclusions drawn from these results and future extensions. 

\section{Background}
\label{sec:background}

\subsection{Wave Equation Model} Assuming a non-lossy medium with homogeneous density and a spatially varying speed of sound (SOS), acoustic propagation in USCT can be modeled using the acoustic wave equation\cite{jensen91}
\begin{equation}\label{eqn:waveq}
\begin{array}{rl}
    \frac{1}{c(\rr)^2}\frac{\partial^2}{\partial t^2} p(\rr,t) - \Delta p(\rr,t) = s(\rr,t)  & (\rr,t) \in \R^d \times (0,T] \\
     \frac{\partial}{\partial t}p(\rr,0) = p(\rr,0) = 0& \rr\in \R^d ,
\end{array}
\end{equation}
where $\sos = \sos(\rr)$ is the spatially varying SOS, $s =s(\rr,t )$ denotes the excitation pulse, $T$ is the data acquisition time, and $p =p(\rr,t )$ is the acoustic pressure field. Eq. \eqref{eqn:waveq} defines a well-posed relationship between the SOS $\sos$, pressure $p$ and source term $s$, which can be expressed as $p = \mathcal{H}^\sos_W s$.

\subsection{Born Approximation} A simplified mathematical model of acoustic wave propagation based on the Born approximation has previously been used in USCT\cite{greenleaf1983computerized, Martiartu20203d, Cardenas2019Strategies, 
stanziola2023learned}. To derive the Born approximation, the pressure field $p = p_i + p_s$ is decomposed into the incident field $p_i$ that satisfies  Eq. \eqref{eqn:waveq} for a fixed constant (or nominal) SOS $c_0$ and the scattered component $p_s$. Under the Born approximation, the approximated scattered component $\tilde{p}_s \approx p_s$ satisfies \cite{greenleaf1983computerized,lebras1988iterative} 

\begin{equation}\label{eqn:scatter_eq}
\begin{array}{rl}
   \frac{1}{c_0(\rr)^2}  \frac{\partial^2}{\partial t}\tilde{p}_s(\rr,t) - \Delta \tilde{p}_s(\rr,t) =    \frac{1}{c_0(\rr)^2} \left(1 - \frac{c_0(\rr)^2}{c(\rr)^2} \right) \frac{\partial^2}{\partial t^2}p_i(\rr,t)  & (\rr,t) \in \R^d \times (0,T] \\
     \frac{\partial}{\partial t}\tilde{p}_s(\rr,0) = \tilde{p}_s(\rr,0) = 0& \rr\in \R^d. 
\end{array}
\end{equation} 
Equation \eqref{eqn:scatter_eq} defines an affine mapping between $\tilde{p} = p_i + \tilde{p}_s \approx p$ and the squared slowness $b(\rr) = \left( \frac{c_0(\rr)}{c(\rr)} \right)^2$. In what follows, the relationship between $\tilde{p}$ and $\sos$ under the Born approximation is denoted as $\tilde{p} = \mathcal{H}^\sos_B s$, where $\tilde{p}$ is linearly dependent on $s$ via its dependence on $p_i$.

When the nominal SOS $c_0$ is constant, the scattered component of the Born approximation can be expressed in a closed-form via the convolution

$$p_s(\rr,t) = \int_{\R^d} \int_0^t G(\rr - \rr', t -t') \left[\frac{1}{c_0^2} \left(1- \frac{c_0^2}{c(\rr')^2} \right) \frac{\partial^2}{\partial {t'}^2} p_i(\rr',t')  \right] dt' d\rr ',$$ where $G$ is the Green's function corresponding to the wave equation for the constant SOS $c_0$.  

Notably, the Born approximation is only sufficiently accurate when the magnitude of the scattered component is much smaller than the incident component \cite{van2006forward}; that is the spatial variations in the SOS are small and the frequency of the excitation source is low. For high SOS contrast media or high-frequency excitation pulses, the use of the Born approximation to reconstruct SOS images from USCT data may lead to severe artifacts due to model mismatch \cite{goncharsky2014inverse, jirik2012sound}.  

\subsection{Continuous-to-Discrete USCT Imaging Operator} Assuming that $J$ idealized point-like transducers are distributed along the measurement aperture $\mathcal{S}$ at locations $\rr_j \in \mathcal{S}$ ($j=1, \ldots J$) and measurements are sampled at $K$ points in time over an acquisition period $[0,T]$, the sampling operator $\mathcal{M}$ mapping the pressure $p(\rr,t )$ to the pressure traces matrix $\boldsymbol{g} \in \mathbb{R}^{K \times J}$ is defined as 
\begin{equation}
 [\mathcal{M} p]_{kj} :=[\boldsymbol{g}]_{kj} = p(\rr_j, k\Delta T), \quad j=1, \ldots J, \, k=1, \ldots K,
\end{equation}
where $\Delta T = T/K$ is the sampling interval. This leads to the continuous-to-discrete imaging relationship 
\begin{equation}
\boldsymbol{g}_i = \mathcal{M}\mathcal{H}^c_W s_i \approx  \mathcal{M}\mathcal{H}^c_{\mathrm{B}} s_i \quad i = 1, \ldots, I,
\end{equation}
where $s_i := s_i(\rr, t)$ is the $i^{\rm th}$ excitation pulse and  $I$ is the number of sources. 

\subsection{Discrete-to-Discrete Imaging Operator}

With the introduction of a Cartesian grid consisting of $Q$ pixels, the imaging operator can be approximated by the corresponding discrete-to-discrete (D-D) imaging operator. Denoting the center of the $q^{\rm th}$ pixel with $\boldsymbol{r}_q$, the finite-dimensional vectors $\boldsymbol{c} \in \mathbb{R}^Q$ and $\boldsymbol{s}_i \in \mathbb{R}^{K Q}$ are defined as 
\begin{equation}\label{eqn:discretization}
    \begin{array}{c}
   [\boldsymbol{c}]_q = c(\rr_q), \; [\boldsymbol{s}_i]_{k+(q-1)K} = s_i(\rr_q, k\Delta T),\ 
     \quad q=1, \ldots, Q; \, k=1,\ldots, K. 
\end{array}
\end{equation} 

With the above notation, the D-D USCT model is given by
\begin{equation}\label{eqn:imaging}
\boldsymbol{g}_i = \boldsymbol{M}\boldsymbol{H}^{\boldsymbol{c}}_W \boldsymbol{s}_i \quad i = 1, \ldots, I,
\end{equation}
where $\boldsymbol{M}: \mathbb{R}^{K Q} \mapsto \mathbb{R}^{K J}$ is the discrete counterpart of the sampling operator $\mathcal{M}$ defined via nearest neighbor interpolation of transducer coordinates to the pixel centers of the Cartesian grid, and $\boldsymbol{H}^{\boldsymbol{c}}_W:  \mathbb{R}^{KQ} \mapsto \mathbb{R}^{K Q}$ stems from finite difference approximation of the wave equation model $\mathcal{H}^\sos_W$. 
The D-D operator $\boldsymbol{H}^{\boldsymbol{c}}_B:  \mathbb{R}^{KQ} \mapsto \mathbb{R}^{K Q}$, stemming from the discretization of  $\mathcal{H}^\sos_B$, is defined in analogous manner.

\subsection{Inversion}

USCT image reconstruction aims at estimating the SOS distribution from noisy transmission and reflection ultrasonic measurements, $\{\data_i\}_{i=1}^I$, defined as
$$\data_i = \boldsymbol{g}_i  + \boldsymbol{n}_i, \quad i = 1, \ldots, I, $$
where $\{\boldsymbol{n}_i\}_{i=1}^I \subset \mathbb{R}^{K \times J}$ denotes additive white noise.

Full waveform inversion~\cite{virieux2009overview} provides an estimate of the SOS distribution by solving an optimization problem of the form
\begin{equation}
\hat{\boldsymbol{c} }_W := \argmin_{\boldsymbol{\sos} \in \R^{{Q}}} \frac{1}{2}\sum_{i=1}^I \left\| \data_i - \boldsymbol{M}\boldsymbol{H}^{\boldsymbol{c}}_W \boldsymbol{s}_i  \right\|^2 + \mathbf{R}(\boldsymbol{c}).
\label{eq:fwi_det}
\end{equation} 
The objective function to be minimized is a weighted sum of a data fidelity term, which evaluating involves the solution of the wave equation for each source, and a convex regularization term, $\mathbf{R}:\mathbb{R}^{Q} \mapsto \mathbb{R}$, which promotes desirable properties of the reconstructed SOS (e.g. smoothness, or edge preservation) and mitigates the ill-posedness of the image reconstruction problem. 

This discrete optimization problem can be solved using a gradient-based method to update estimates of $\boldsymbol{\sos}$. However, computing the gradient of the objective function with respect to $\boldsymbol{\sos}$ requires the solution to $I$ different forward and adjoint wave equations, and thereby entails large computational cost. To drastically reduce the computational cost, this work utilizes a stochastic  \emph{source encoding} approach, which leverages the linearity of the imaging operator $\boldsymbol{H}^{\boldsymbol{sos}}_W$ with respect to the excitation pulse $\boldsymbol{s}$ \cite{krebs09, wang2015waveform, lucka21, bachmann2020source}. Source encoding reformulates the original optimization problem in \eqref{eq:fwi_det} as the stochastic optimization problem 

\begin{equation}
\hat{\boldsymbol{c} }_W := \argmin_{\boldsymbol{\sos} \in \R^Q} \frac{1}{2}  \mathop{\mathbb{E}}_{\boldsymbol{a} \sim \Gamma} \left\| \data_{\boldsymbol{a}} -  \boldsymbol{M}\boldsymbol{H}^{\boldsymbol{c}}_W \boldsymbol{s}_{\boldsymbol{a}} \right\|^2 + \mathbf{R}(\boldsymbol{c}),
\label{eq:wise}
\end{equation} 
where $\boldsymbol{a} \in \mathbb{R}^I$ is the stochastic encoding vector sampled according to a multivariate probability distribution $\Gamma$ with zero mean and identity covariance matrix (such as a Rademacher or standard normal distribution), and $\data_{\boldsymbol{a}} = \sum_{i=1}^I [\boldsymbol{a}]_i \data_i$, $\boldsymbol{s}_{\boldsymbol{a}} = \sum_{i=1}^I [\boldsymbol{a}]_i \boldsymbol{s}_i$ denote the superimposed (encoded) measurement data and excitation source, respectively. Using stochastic gradient descent to find an approximate minimizer of Eq. \eqref{eq:wise} requires only one forward and backward wave equation solution at each iteration, compared to the $I$ solutions required to evaluate the gradient of Eq. \eqref{eq:fwi_det}. However, even when advanced acceleration techniques are employed \cite{nesterov1983method, polyak1964some, KingmaBa2014}, convergence remains slow due to the nonconvex nature of the optimization problem and a single reconstruction incurs significant computational burden (several minutes or hours).

Under the Born approximation, an estimate of the SOS distribution can be obtained by solving a modified optimization problem of the form and Eq. \eqref{eq:wise} yielding
\begin{equation}
\hat{\boldsymbol{c} }_B := \argmin_{\boldsymbol{\sos} \in \R^Q} \frac{1}{2}\sum_{i=1}^I \left\| \data_i - \boldsymbol{M}\boldsymbol{H}^{\boldsymbol{c}}_B \boldsymbol{s}_i  \right\|^2 + \mathbf{R}(\boldsymbol{c}),
\label{eq:born_det}
\end{equation}
where $\boldsymbol{H}^{\boldsymbol{c}}_W \boldsymbol{s}$ in Eqs.~\eqref{eq:fwi_det} is replaced by $\boldsymbol{H}^{\boldsymbol{c}}_B \boldsymbol{s}$. 

Similarly to Eq. \eqref{eq:wise}, the use of source encoding leads to the following stochastic optimization form of the Born inversion approach:
\begin{equation}
\hat{\boldsymbol{c} }_B := \argmin_{\boldsymbol{\sos} \in \R^Q} \frac{1}{2}  \mathop{\mathbb{E}}_{\boldsymbol{a} \sim \Gamma}\left\| \data_{\boldsymbol{a}} -  \boldsymbol{M}\boldsymbol{H}^{\boldsymbol{c}}_B \boldsymbol{s}_{\boldsymbol{a}} \right\|^2 + \mathbf{R}(\boldsymbol{c}).
\label{eq:born_wise}
\end{equation}

Notably, the optimization problems in Eqs.~\eqref{eq:born_det} and Eq. \eqref{eq:born_wise} 
are in the form of a penalized linear least-squares problem for the squared slowness $\boldsymbol{b} = \left( \frac{\boldsymbol{c}_0}{\boldsymbol{c}}\right)^2$. Their solution incurs reduced computational expense; however, modeling errors introduced by the use of the Born approximation may hinder accuracy and resolution in the reconstructed image.

\subsection{Task-Based Assessment of Image Quality}

Medical imaging is usually performed for some screening, diagnostic, or therapeutic task. However, physical metrics of image quality, such as root mean square error or structural similarity, do not always directly correlate with the usefulness of a reconstructed image for performing a task of interest\cite{barrett93, christianson15}. In fact, some recent studies have demonstrated that, while learning-based approaches can often improve perceptual and physical measurement of image quality, they may also introduce subtle hallucinations\cite{bhadra21hallucinations} and reduce clinical utility (task performance)\cite{li21}. Due to this discrepancy, there is a growing interest in assessing the task performance, via a task-based measure of image quality, of proposed image reconstruction methods alongside physical metrics of image accuracy \cite{zhang21, adler22, he2013model}. In these assessments, numerical observers \cite{barrett93,barrett2013foundations} are often employed to perform the task of interest on a large ensemble of reconstructed images from the proposed reconstruction method, and the area under the receiver-operating characteristic (ROC) curve is often employed as a figure of merit for task performance.

In USCT breast imaging, a clinical task of interest is tumor detection and localization, or segmentation.  An image reconstruction method for USCT can then be assessed by an observer's performance for tumor segmentation on the reconstructed images. Notably, in this task performance assessment, a wide variety of medically realistic example images are utilized. Simulating reconstruction in a wide variety of examples is important to understand the efficacy of a USCT method's  task performance across breasts with different sizes and densities and ensure equivalent performance across populations \cite{PhantomGen}.

\section{Method}\label{sec:method}
This section presents the main contribution of this work: a systematic comparison of machine learning methods for modeling error compensation in USCT breast imaging. Although the primary focus of this work is USCT breast imaging, the ideas explored are generalizable to a wide variety of applications for which models with varying fidelity are available. 

Specifically, this work performs a systematic analysis of three hybrid approaches for USCT image reconstruction utilizing the Born approximation and a purely data-driven learned reconstruction method based on their quality with respect to physical metrics of image quality and task-based assessment of image quality. This assessment utilizes the FWI method and an uncorrected Born inversion method as reference methods. These hybrid approaches employ learned correction before or after reconstruction, or a combination of both. The first hybrid approach, \emph{artifact correction}, utilizes the simplified model for reconstruction and then applies a neural network to correct for artifacts due to model mismatch. The second hybrid approach, \emph{data correction}, utilizes a neural network to preprocess, or correct, measurements and then invert utilizing a simplified physical model. The third hybrid approach, \emph{dual correction}, is a dual method that utilizes one neural network to preprocess data, invert with the Born approximation, and then use a second network to correct the reconstructed image. Additionally, these hybrid approaches are compared to a purely data-driven learned reconstruction method, \emph{InversionNet}, which utilizes a neural network to map from measurements to a reconstructed image\cite{inversionNET}.

\subsection{Artifact Correction}\label{subsec:artifact_correction}

Let $\Phi_\eta: \R^{Q}\rightarrow \R^Q$ denote a CNN with weights $\eta\in \R^{W_{\mathrm{AC}}}$ that maps from a set of  inaccurate reconstructions based on an approximated imaging model to a set of accurate reconstructions with features informed by  a set of training objects. 
This network is trained in a supervised manner by minimizing the empirical minimum square error loss

$$\min_\eta \sum_{n=1}^N \| \Phi_\eta(\hat{\boldsymbol{\sos}}_{\mathrm{B}}^n) - \boldsymbol{\sos}^n\|^2,$$ for a set of training SOS maps $\{\boldsymbol{\sos}^n\}_{n=1}^N$ and Born reconstructions $\{\hat{\boldsymbol{\sos}}_{\mathrm{B}}^n\}_{n=1}^N$, as defined in Eqs. \eqref{eq:born_det} and \eqref{eq:born_wise}. Note that the training set reconstructions $\{\hat{\boldsymbol{\sos}}_{\mathrm{B}}^n\}_{n=1}^N$ depend on the choice and strength of the regularization $\mathbf{R}$ in Eq. \eqref{eq:born_det} and Eq. \eqref{eq:born_wise}.  This dependence on regularization means that an artifact correction method needs to be retrained for a new choice of regularization and each instance can only be applied to a single reconstruction approach.

Once trained, this network can be used for reconstruction by applying the neural network to the solution of an optimization problem, 

$$\hat{\boldsymbol{c}}_{\mathrm{AC}} := \Phi_\eta \left( \argmin_{\boldsymbol{\sos} \in \R^Q} \frac{1}{2}\sum_{i=1}^I \left\|\data_i - \boldsymbol{M}\boldsymbol{H}^{\boldsymbol{c}}_{\mathrm{B}} \boldsymbol{s}_i\right\|^2  + \boldsymbol{R}(\boldsymbol \sos) \right),$$ or equivalently utilizing source encoding

$$\hat{\boldsymbol{c}}_{\mathrm{AC}} := \Phi_\eta \left( \argmin_{\boldsymbol{\sos} \in \R^Q} \frac{1}{2}  \mathop{\mathbb{E}}_{\boldsymbol{a} \sim \Gamma}\left\| \data_{\boldsymbol{a}} -  \boldsymbol{M}\boldsymbol{H}^{\boldsymbol{c}}_B \boldsymbol{s}_{\boldsymbol{a}} \right\|^2 + \mathbf{R}(\boldsymbol{c}) \right).$$

\subsection{Measurement Correction}\label{subsec:measurement_correction}

Let $\Psi_\xi:\R^{KJ} \rightarrow \R^{KJ}$ denote a CNN with weights $\xi \in \R^{W_{\mathrm{DC}}}$ that maps data modeled by the wave equation to data modeled by the Born approximation. 
This network is trained in a supervised manner by minimizing the data-domain empirical minimum square error loss

\begin{equation}\label{eqn:data_correction}
    \min_\xi \frac{1}{2} \sum_{n=1}^N\sum_{i=1}^I\| \boldsymbol{\mathfrak{d}}_i^n - \Psi_\xi( \boldsymbol{d}_i^n ) \|^2,  
\end{equation} with $\boldsymbol{d}_i^n$ defined in Eq. \eqref{eqn:imaging} and 

\begin{equation} \boldsymbol{\mathfrak{d}}_i^n \coloneqq \boldsymbol{M}\boldsymbol{H}^{\boldsymbol{c}^n}_{\mathrm{B}} \boldsymbol{s}_i,
\end{equation} for a set of training SOS maps $\{ \boldsymbol{c}^n \}_{n=1}^N$.  Once trained, this network can be used to preprocess the data before solving the linear penalized least-square problem
$$\hat{\boldsymbol{\sos}}_{\mathrm{DC}} :=\argmin_{\boldsymbol{\sos} \in \R^Q} \frac{1}{2}\sum_{i=1}^I\|\Psi_\xi(\data_i) - \boldsymbol{M}\boldsymbol{H}^{\boldsymbol{c}}_{\mathrm{B}} \boldsymbol{s}_i\|^2 + \boldsymbol{R}(\boldsymbol{\sos}),$$  or equivalently utilizing source encoding

\begin{equation}
\hat{\boldsymbol{\sos}}_{\mathrm{DC}} := \argmin_{\boldsymbol{\sos} \in \R^Q} \frac{1}{2}  \mathop{\mathbb{E}}_{\boldsymbol{a} \sim \Gamma}\left\| \sum_{i=1}^I [\boldsymbol{a}]_i\Psi_\xi(\data_{i}) -  \boldsymbol{M}\boldsymbol{H}^{\boldsymbol{c}}_B \boldsymbol{s}_{\boldsymbol{a}} \right\|^2 + \mathbf{R}(\boldsymbol{c}).
\label{eq:data_corrected_sos_encoding}
\end{equation}

Note that training the process for the network $\Psi_\xi$ does not depend on the choice of regularization term $ \mathbf{R}$. Instead, this measurement correction approach can be inserted into any choice of a reconstruction algorithm that utilizes the Born approximation. 

\subsection{Dual Correction}\label{subsec:dual_correction}

Approaches utilizing both measurement~(or data) correction and artifact correction can be combined into a dual correction formulation. The goal of such a dual correction method is to capitalize on the perceived benefits from both the data correction and artifact correction methods. Specifically, a dual-corrected method may present less bias than an artifact-corrected method and higher image quality than a measurement-corrected method.

Let $\Phi_\eta: \R^{Q}\rightarrow \R^Q$ denote a CNN with weights $\eta\in \R^{W_{\mathrm{AC}}}$ designed for artifact correction and let $\Psi_\xi:\R^{KJ} \rightarrow \R^{KJ}$ be a CNN with weights $\xi \in \R^{W_{\mathrm{DC}}}$ designed for data correction. The network $\Psi_\xi$ previously trained in equation Eq. \eqref{eqn:data_correction} can be reused in data correction; however, the artifact correction network $\Phi_\eta$ needs to be retrained because the input images are generated using $\Psi_\xi(\data)$ rather than $\data$. This network is trained in a supervised manner by minimizing the image-domain mean square error loss

$$\min_\eta \sum_{n=1}^N \| \Phi_\eta(\hat{\boldsymbol{\sos}}_{\mathrm{DC}}^n) - \boldsymbol{\sos}^n\|^2,$$
where $\hat{\boldsymbol{\sos}}_{\mathrm{DC}}$ is defined by Eq. \eqref{eq:data_corrected_sos_encoding}.

Once trained, these networks can be used for inversion by

$$\hat{\boldsymbol{c}}_{\mathrm{Dual}} := \Phi_\eta \left( \argmin_{\boldsymbol{\sos} \in \R^Q} \frac{1}{2}\sum_{i=1}^I \left\|\Psi_\xi(\data_i) - \boldsymbol{M}\boldsymbol{H}^{\boldsymbol{c}}_{\mathrm{B}} \boldsymbol{s}_i\right\|^2  + \boldsymbol{R}(\boldsymbol{\sos}) \right) \;,$$ 
or equivalently utilizing source encoding 

$$\hat{\boldsymbol{c}}_{\mathrm{Dual}} := \Phi_\eta \left( \argmin_{\boldsymbol{\sos} \in \R^Q} \frac{1}{2}  \mathop{\mathbb{E}}_{\boldsymbol{a} \sim \Gamma}\left\| \sum_{i=1}^I [\boldsymbol{a}]_i\Psi_\xi(\data_{i}) -  \boldsymbol{M}\boldsymbol{H}^{\boldsymbol{c}}_B \boldsymbol{s}_{\boldsymbol{a}} \right\|^2 + \mathbf{R}(\boldsymbol{c}) \right).$$

\subsection{Data-Driven Learned Reconstruction}\label{subsec:data_driven}

Previous works that have applied machine learning for USCT image reconstruction have proposed fully learned reconstruction methods that utilize a CNN to learn a mapping
from measurements to reconstructed images \cite{lozenski2024usct, fan21, inversionNET, jin2022unsupervised, feng2022intriguing}. 

Let $\Theta_\nu: \R^{KJ\times I} \rightarrow \R^{Q}$ denote a CNN with weights $\nu \in \R^P$ that maps pressure waveforms to the reconstructed SOS maps. Such a network can be trained in a supervised manner by minimizing the minimum square error loss 

$$\min_\nu \frac{1}{2} \sum_{n=1}^N\| \Theta_\nu (\data_1^n , \hdots, \data_I^n) - \boldsymbol{\sos}^n \|^2,$$ for a set of training SOS maps $\{\boldsymbol{\sos}^n\}_{n=1}^N.$ Once trained, the evaluation of this network can be directly utilized for the reconstruction 

$$\hat{\boldsymbol{\sos}}_{DD} := \Theta_\nu (\data_1,\hdots,\data_I).$$

\subsection{Task-Based Assessment of Image Quality}

Aside from physical metrics of image quality, such as mean square error or structural similarity index measure \cite{WangBoviketall04}, this work seeks to include a task-based assessment of image quality utilizing machine learning tools.  

The task assessed in this work was tumor detection and localization by use of a numerical observer. Specifically, this work utilized a convolutional neural network to evaluate the numerical observer\cite{pereira2016brain, havaei2017brain, he2013model}. The neural network-based observer is denote as $\Xi_\omega:\R^Q \rightarrow [0,1]^Q$, and parameterized by $\omega \in \R^{W_{O}}$. It acts as a map from discrete SOS images $\boldsymbol{\sos}$ to pointwise probability maps $\hat{\tumorlabelmap} := \Xi_\omega({\boldsymbol{\sos}})$. Here, the $q$-th entry of $\hat{\tumorlabelmap}$ represents the probability that the $q$-th pixel belongs to a tumor region. Such a network can be trained in a supervised manner by solving the minimization problem 

$$\min_{\omega} \sum_{n=1}^N H({\tumorlabelmap}^n, \Xi_\omega({\boldsymbol{\sos}}^n)), $$
where $H$ is a loss function such as pixel-wise cross entropy \cite{de2005tutorial}, $\{{\boldsymbol{\sos}}^n\}_{n=1}^N$  is a set of SOS training maps, and $\{\tumorlabelmap^n\}_{n=1}^N$ is a set of tumor segmentation masks (labels). 

Once trained, a receiver operator characteristic (ROC) analysis \cite{zweig1993receiver} of the learned numerical observer can be performed on a set of testing images. In this analysis, a ROC curve is constructed by sweeping a threshold probability $\tau$ and plotting the resulting parametric curve with the false detection rate $P([\Xi_\omega({\boldsymbol{\sos}})]_q > \tau \,| \, [{\tumorlabelmap}]_q = 0)$ on the $x$-axis and the true detection rate $P([\Xi_\omega({\boldsymbol{\sos}})]_q > \tau \,|\, {[\tumorlabelmap}]_q = 1)$ on the $y$-axis. The resulting ROC curve illustrates the observer's performance and the area under the curve (AUC) serves as a quantitative metric for overall task performance.

For each reconstruction method, an observer can be trained on the training set reconstructions. The resulting ROC curve and AUC can be used to assess and quantify a reconstruction method's ability to preserve task-relevant information and the resulting images' usefulness for performing the tumor detection task. Similarly,  an observer can be trained on the set of true training objects. The  AUC corresponding to the observer trained on the true objects serves as an upper bound for the AUC of the other observers and demonstrates the feasibility of learning the task from the training set objects and transferring it to the testing set.  

\begin{figure*}[tbh]
    \centering\includegraphics[width = \textwidth]{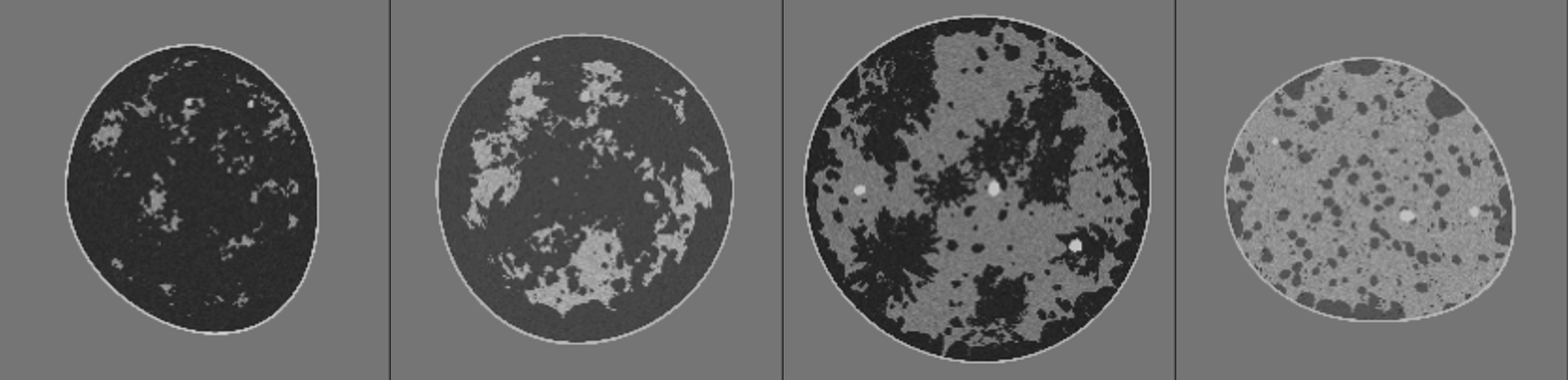}
    \caption{Four examples of the anatomically realistic numerical breast phantoms (NBPs), one from each of BI-RADS breast density types, used to train and evaluate the proposed learned correction methods. NBPs present clinically relevant variability in size, tissue composition and structures, and SOS maps. From left to right: Type A (almost entirely fatty), Type B (scattered fibroglandular density), Type C (heterogeneous density), and Type D (extremely dense). }
    \label{fig:examples_NBPs}
\end{figure*}

\section{Numerical Studies}\label{sec:numerical_studies} \subsection{Construction of the numerical phantoms}

The numerical studies presented in this work made use of anatomically realistic numerical breast phantoms (NBPs) to which spatially varying SOS values were stochastically assigned within feasible ranges. These NBPs were developed by Li et al. \cite{PhantomGen} for use in USCT virtual imaging studies using tools adapted from the Virtual Imaging Clinical Trial for Regulatory Evaluation (VICTRE) project at the US Food and Drugs Administration \cite{badano2018evaluation}. Examples of these NBPs are available from \cite{LiDataverse2021}. In particular, the  generated NBPs are stratified based on the four different levels of breast density (percentage of fibroglandular tissue) defined according to the American College of Radiology's (ACR) Breast Imaging Reporting and Data System (BI-RADS) \cite{american2013acr}: (A) almost entirely fatty breasts, (B) breasts with scattered fibroglandular density, (C) breasts with heterogeneous density, and (D) extremely dense breasts.  Four realizations of these stochastic NBPs, one from each of the BI-RADs categories, are shown in Fig. \ref{fig:examples_NBPs}. 

\begin{table}[tb]
\vspace{-0mm}
\caption{Numerical study parameters}
    \centering
    \begin{tabular}{ll}
    \hline
    \multicolumn{2}{c}{Virtual USCT system}\\
    \hline
        Number of sources $I$ & 64 \\
        Number of receivers $M$ & 256 \\
        Transducer radius $R$ & 96 mm \\ 
        Time horizon $T$ & 160 $\mu s$ \\ 
        Sampling frequency & 5 MHz \\ 
        Pulse central frequency $f_0$ & $500$ kHz \\ 
        Pulse center $t_0$ & 3.2 $\mu$s\\ 
        Pulse width $\sigma$ & 10 $\mu$s \\ 
        \hline
    \end{tabular}
\hspace{1cm}
\begin{tabular}{ll}
        \hline
        \multicolumn{2}{c}{ Discretization details}\\
        \hline

                Grid size $N_x$ & 360 \\ 
                Field of view size $n_x$ & 214 \\
        Grid interval $\delta x$ & 0.6 mm \\
        Number of time steps $K$ & 800 \\ 
        Time step interval $\delta t$ & 0.2 $\mu$s\\ 
        CFL Number  $\frac{\sos_{max} \delta t}{\delta x}$ & 0.53 \\ 
        Points per wavelength  $\frac{\sos_{min}}{f_0 \delta x}$ & 4.7\\ 
        \hline
        \phantom{ fix align} & \phantom{ fix align}\\
    \end{tabular}
    \label{tab:ultrasound_system}
\end{table}

\subsection{Definition of the virtual imaging systems}
\label{subsec:imaging_system}

The numerical studies in this work considered an idealized system motivated by clinical ring-array breast USCT systems. The measurements geometry consisted of a circular transducer array\cite{wang2015waveform} of radius $R=96\, {\rm mm}$ along which 256 idealized point-like transducers were equispacially distributed.  Each of the 256 transducers acted as receivers. Every fourth transducer, 64 in total, also acted as a transmitter and would emit an excitation pulse in sequence. The $i$-th excitation pulse was of the form
$$\begin{array}{c}
     s_i(\rr,t) = \delta(\rr - \rr_i)\exp\left( -\frac{(t-t_0)^2}{2\sigma^2} \right) \sin(2\pi f_0 t), \
     i = 1,\hdots, I,
\end{array}$$
where $\rr_i$ is the location of the $i$-th emitter, $f_0 = 500\,{\rm kHz}$ is the central frequency, $t_0=3.2\,\mu$s is the time shift, and $\sigma=2\,\mu$s controls the signal width. Measurements were collected by firing one transmitter at a time and recording data at every receiver. Waveform data generated by each source were collected over an acquisition time $T=160 \, \mu{\rm s}$, which is long enough to capture secondary wavefront arrivals. This is repeated for each emitter and results in multi-channel measurements.

To numerically simulate the pressure field generated by each emitter, the wave equation in Eq. \eqref{eqn:waveq} was 
solved using a finite difference scheme (4th order in space and 2nd order in time) with an absorbing boundary condition to mitigate wave reflections introduced by the finite extension of the computational domain. A spatial grid of size $N_x\times N_x$, with $N_x=360$, and a temporal grid with $K = 800$ samples were employed for the discretization. Electronic noise was modeled as additive white Gaussian noise. 
Reconstructions were  performed on a field of view corresponding to a spatial grid with $n_x\times n_x$ pixels, where $n_x = 214$ and  the same pixel sizes  and time steps as utilized in the forward simulation. The imaging system parameters are reported in Table \ref{tab:ultrasound_system}.

\subsection{Study Design}

Simulation studies were performed to compare the viability and accuracy of the four methods presented in Section \ref{sec:method}. These studies considered six different reconstruction methods applied to the waveform data generated by solving the wave equation model in Eq. \eqref{eqn:waveq}. The first method utilized the FWI method, where the exact imaging operator was used for inversion. The second method utilized a large CNN, with the \emph{InversionNet} architecture \cite{inversionNET}, as a representative data-driven learned reconstruction method as described in Subsection \ref{subsec:data_driven}. The third method (\emph{Uncorrected}) inverted the waveform data using the Born approximation as the imaging operator. The fourth method (\emph{Artifact Corrected}) inverted waveform data using the Born approximation, then utilized a CNN to correct artifacts due to model mismatch as described in Subsection \ref{subsec:artifact_correction}. The fifth method (\emph{Data Corrected}) utilized a CNN for learned measurement correction and then inverted using the Born approximation as described in Subsection \ref{subsec:measurement_correction}.  The sixth method (\emph{Dual Corrected}) applied the CNN for measurement correction, inverted using the Born approximation, then applied a second CNN for artifact correction as described in Subsection \ref{subsec:dual_correction}.

 The first study considered the \emph{in-distribution} assessment of the methods, in which both the training and testing sets were comprised of all types of breast phantoms. The second study considered \emph{out-of-distribution} assessment, in which the training set consisted of fatty to moderately dense breasts (types A, B, C in Fig. \ref{fig:examples_NBPs}), while the testing set only contained extremely dense breast (type D). The third study assessed the effect of the number of training examples on the generalizability and accuracy of the methods considered. The fourth study assessed the \emph{robustness} to measurement noise of each learned reconstruction method. 

\subsubsection{Study 1: In-distribution assessment}

The first numerical study considered \emph{in-distribution} assessment of image reconstruction. In this study, the training and testing sets were both generated from the full distribution of breast phantoms corresponding to all BI-RADS categories.  This study serves as the baseline for the other studies and represents ideal conditions for machine learning methods, in which the training and testing set are sampled from the same distribution. In this study,  the training set consisted of 820 NBPs featuring 88 Type A NBPs, 274 Type B NBPs, 272 Type C NBPs, and 186 Type D NBPs. The testing set consisted of 615 NBPs featuring 67 Type A NBPs, 206 Type B NBPs, 204 Type C NBPs, and 138 Type D NBPs.  The measurement noise had a standard deviation of $3.0 \times 10^{-5},$ corresponding to a signal-to-noise ratio (SNR) of 20 dB.

\subsubsection{Study 2: Out-of-distribution assessment}

 The second study considered \emph{out-of-distribution} assessment of image reconstruction. In this study, the training set only included the low density to moderately dense breasts, corresponding to BI-RADS Types A-C, while the testing set only contained extremely dense breasts, corresponding to BI-RADS Type D. 

 In breast cancer screenings, mammograms are less sensitive to detecting tumors in Type D breasts but USCT is sensitive to detecting tumors in Type D breasts.  \cite{duric2005development}. However, USCT image reconstruction for Type D breasts can still be more challenging compared to other breast types due to higher heterogeneity in SOS \cite{PhantomGen}. Additionally, Type D breasts are an underrepresented population, corresponding to ten percent of mammograms performed \cite{sprague2014prevalence}. Type D breasts are, therefore, an important holdout set to assess generalizability to out-of-distribution data. Thus, this study serves to assess the robustness of the considered approaches with respect to a domain shift (i.e., training on fattier breast data and testing on denser breast data).
 
In this study, the training set consisted of  1,134 NBPs featuring 162 Type A NBPs, 488 Type B NBPs, and 484 Type C NBPs. The testing set consisted of 324 type D NBPs.  As in the first study, the measurement noise had a standard deviation of $3.0 \times 10^{-5},$ corresponding to an SNR of 20 dB.

\subsubsection{Study 3: Effect of  training data paucity}

The third study assessed the image reconstruction methods when the size of the training set was limited. In this study, all learning methods were trained using a training set that was half of the size of the one used in Study 1. The methods were then assessed on all other NBPs, resulting in a testing set that was effectively larger than that used in Study 1.

This study assessed in-distribution reconstruction with half the training examples used in the first study and the other half moved to the testing set. This demonstrates each reconstruction method's ability to learn from limited training data. In this study,  the training set consisted of 410 NBPs featuring 44 Type A NBPs, 137 Type B NBPs, 136 Type C NBPs, and 93 Type D NBPs.  The testing set consisted of 1,025 NBPs featuring 111 Type A NBPs, 343 Type B NBPs, 340 Type C NBPs, and 231 Type D NBPs.  As in the previous studies, the measurement noise had a standard deviation of $3.0 \times 10^{-5},$ corresponding to an SNR of 20 dB.

\subsubsection{Study 4:  Robustness with respect to measurement noise}

 The fourth study assessed each image reconstruction method's  \emph{robustness} with respect to multiple levels of measurement noise in training and testing noise.
 This study repeated the training process from Study 1 using higher levels of measurement noise (medium and high), using the same training and testing splits. The medium level of measurement noise had a standard deviation of $6.0 \times 10^{-5},$ corresponding to an SNR of 14 dB. The high level of measurement noise had a standard deviation of $1.5 \times 10^{-4},$ corresponding to an SNR of 6 dB.  Once these additional methods were trained, they were evaluated, along with the low noise cases, on all three noise levels, and the effect of changing noise levels and mismatch between training and was evaluated.

\subsection{Image Quality Assessment Criteria}

Image quality was assessed using three metrics. The first metric was the relative root mean square error (RRMSE) defined as 
$$ \textnormal{RRMSE}(\hat{\boldsymbol{\sos}}) := \frac{\|\boldsymbol{\sos}_{\mathrm{true}}- \hat{\boldsymbol{\sos}}\|_2}{\|\boldsymbol{\sos}_{\mathrm{true}}- \boldsymbol{\sos}_{\mathrm{bg}}\|_2},$$
where $\hat{\boldsymbol{\sos}}$ represents the estimated SOS map, $\boldsymbol{\sos}_{true}$ represents the object SOS distribution, and $\boldsymbol{\sos}_{\mathrm{bg}}$ represents a given background SOS (water).  The second metric was the structural similarity index measure (SSIM) as defined in Wang et al. \cite{WangBoviketall04}. This serves as a metric of perceptual image quality.  The third metric was a task-based measurement of image quality. In each study a numerical observer, utilizing a U-Net architecture \cite{ronneberger2015u}, was trained to segment the tumor region in the estimated SOS images reconstructed using each method. The same training set was used for the training of both the ML-based image reconstruction methods and the numerical observer network. An ROC analysis of the pixel-wise detection task was then performed utilizing the reconstructed SOS images from the testing set. The AUC under the ROC served as a quantitative metric of task performance. For each study, the numerical observer was also trained utilizing the ground truth training objects  and ROC analysis was performed on the ground truth testing images to define a performance upper bound for any numerical observer acting on reconstructed SOS maps.

\section{Results}\label{sec:results}

Image reconstruction and network training were performed on an HPC node with 512 GB of memory and an NVidia Volta V100 graphic processing unit (GPU). Each neural network corresponding to the methods described in Sect \ref{sec:method} was trained utilizing the same computational budget,  approximately 10 GPU hours. The networks utilized for both data and artifact correction utilized a U-Net ~\cite{ronneberger2015u} architecture with 1,127,859 trainable parameters. The data correction network was trained for approximately 2,000 training epochs while the artifact correction networks were trained for approximately 50,000 epochs. The learned reconstruction method with an InversionNet architecture utilized 117,762,947 parameters and was trained for approximately 1,000 epochs. Visual inspection of training and validation loss curves as a function of the epoch number indicated that all methods were trained for a sufficiently long number of epochs.

For each study, reconstructing one image in the testing set utilizing the \emph{FWI} method required approximately 8.4 GPU minutes. Comparatively, reconstructing one image using the \emph{Uncorrected}, \emph{Artifact Corrected}, \emph{Data Corrected}, and \emph{Dual Corrected} methods required approximately 50 GPU seconds, each. Notably, most of the computational time was spent in performing the Born inversion and the application of the data and artifact correction only required approximately 0.05 GPU seconds for each image. The data-driven reconstructions utilizing \emph{InversionNet} were much faster and required approximately 0.5 GPU seconds for each image, corresponding to the time for one forward pass of the neural network. 
 
\subsection{Study 1: In-distribution assessment}

\subsubsection{Qualitative assessment}
The top row of Fig. \ref{fig:example_in} shows an example of SOS representative of a Type D breast along with the SOS reconstructed estimates produced by each method. The middle row of the figure showed a zoomed-in inset where two lesion are present. The bottom row of the figure shows the lesion segmentation map of the object and the thresholded output of the tumor detection/localization numerical observer applied to each reconstructed image. 
The \emph{FWI} reconstruction demonstrates little or no discrepancy from the true object. The \emph{InversionNet} reconstruction is very inaccurate and poorly captures internal tissue structures or lesions. The \emph{Uncorrected} reconstruction presents strong artifacts caused by model mismatch that obscures the tumor features present. Applying \emph{Artifact Correction} results in a hallucinated image that seems plausible and high resolution but does not preserve detectable lesions present in the object.  The \emph{Data Corrected} reconstruction has minor artifacts due to model mismatch, but can lead to accurate tumor segmentation.  Finally, the \emph{Dual corrected} method lead to reconstructed images with the highest visual similarity to true object among all learned methods; however, the corresponding tumor segmentation map  is less accurate compared to one produced using the SOS map estimated by the \emph{Data Corrected} method.

\subsubsection{Quantitative assessment}
\begin{figure}
        \centering
        \includegraphics[width = 0.63\textwidth, trim = {0cm 0cm 8cm 1.5cm}, clip]{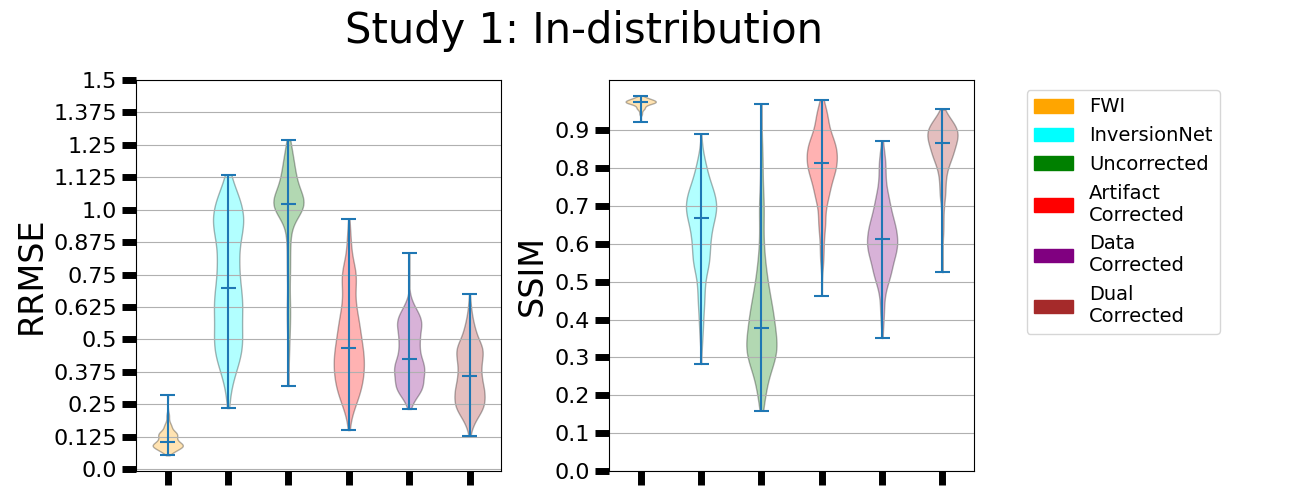}\includegraphics[width = 0.36\textwidth,  trim = {0cm 0cm 0cm 1.32cm}, clip]{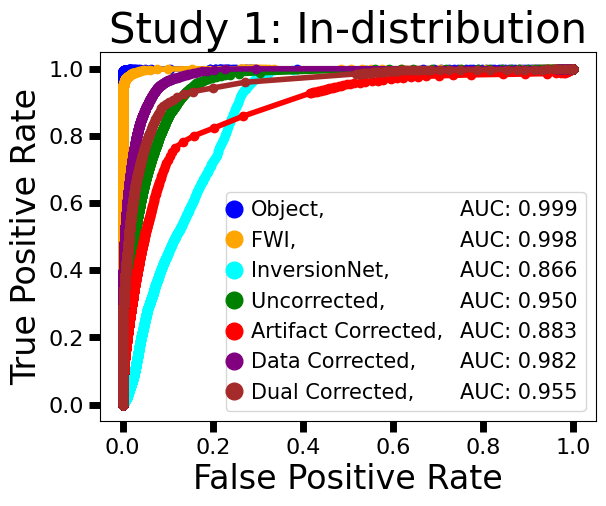}
        \caption{Study 1: In-distribution assessment. Left and middle panels: Violin plots for relative root mean square error (RRMSE) and structural similarity index (SSIM) achieved by each reconstruction method on the testing set data. Right panel: receiver operator characteristic (ROC) curve for pixel-wise tumor detection task computed using reconstructed images from the  testing set. A numerical observer was trained independently for each reconstruction method. Notably, the \emph{FWI} method, which serves as reference, greatly outperforms the other methods in terms of all three metrics.  The \emph{Dual Corrected} method demonstrates the second best performance in terms of RRMSE and SSIM. The \emph{Data Corrected} method performs better than the \emph{Artifact Corrected} and \emph{InversionNet} methods in terms of RRMSE but worse in terms of SSIM. Notably, the \emph{Artifact Corrected} method demonstrates a decrease in task performance, as demonstrated by the area under the ROC curve, when compared to the \emph{Uncorrected} method, with a similar decrease in performance between the \emph{Dual Corrected} and \emph{Data Corrected} methods. }
        \label{fig:violin_plots_in}
    \end{figure}

The ensemble RRMSE and SSIM violin plots and ROC curves computed on images reconstructed by each reconstruction method using data from the testing set are displayed in Fig. \ref{fig:violin_plots_in}. These plots display the distribution of each sample population distributed along the y-axis and bars for the min, max, and median of each population. Among all learned reconstruction methods, the \emph{Dual Corrected} method achieved the best performance in terms of RRMSE and SSIM while the \emph{Data Corrected} method resulted in the best task performance out of all learning methods, as illustrated by the AUCs of the ROC curves. Unsurprisingly, the FWI method statistically outperformed all  methods in terms of RRMSE and SSIM (p-values $<10^{-80}$) and task performance as demonstrated by AUC. However, the improvement utilizing the \emph{Dual Corrected} methods was also statistically significant compared to the \emph{Uncorrected, Artifact Corrected,} and \emph{Data Corrected} methods (p-values  $< 10^{-6}$).

Furthermore, the methods employing artifact correction in the image domain, \emph{Artifact Corrected} and \emph{Dual Corrected}, demonstrated reduced task performance compared to their counterparts without artifact correction, \emph{Uncorrected} and \emph{Data Corrected}, respectively, suggesting that artifact correction removes relevant features. Finally, the \emph{Data Corrected} method led to improved task performance over the \emph{Uncorrected} method, suggesting data correction enhances task-relevant features.

\subsubsection{Discussion}

This study highlights the advantages and disadvantages of approaches employing data correction in the measurement domain compared to artifact correction in the image domain. Employing data correction leads to reconstructed images with higher accuracy in terms of RRMSE while preserving task-relevant information, but lacks perceptive quality as described by SSIM. Comparatively, artifact correction seems to result in plausible images with high perceptive quality; however, these images lack quantitative accuracy and may exhibit hallucinated features and loss of task-relevant information \cite{zhang21}. Combining both forms of correction, in a dual correction method, results in best results in terms of image accuracy and perceptive quality, with a slight reduction in task performance. Furthermore, these approaches incorporating an approximated physical model greatly outperform a purely data-driven method that does not incorporate any physical model. 

    \begin{figure}
        \centering
        \includegraphics[width = \textwidth, trim = {0cm 0.5cm 0cm 0.7cm}, clip]{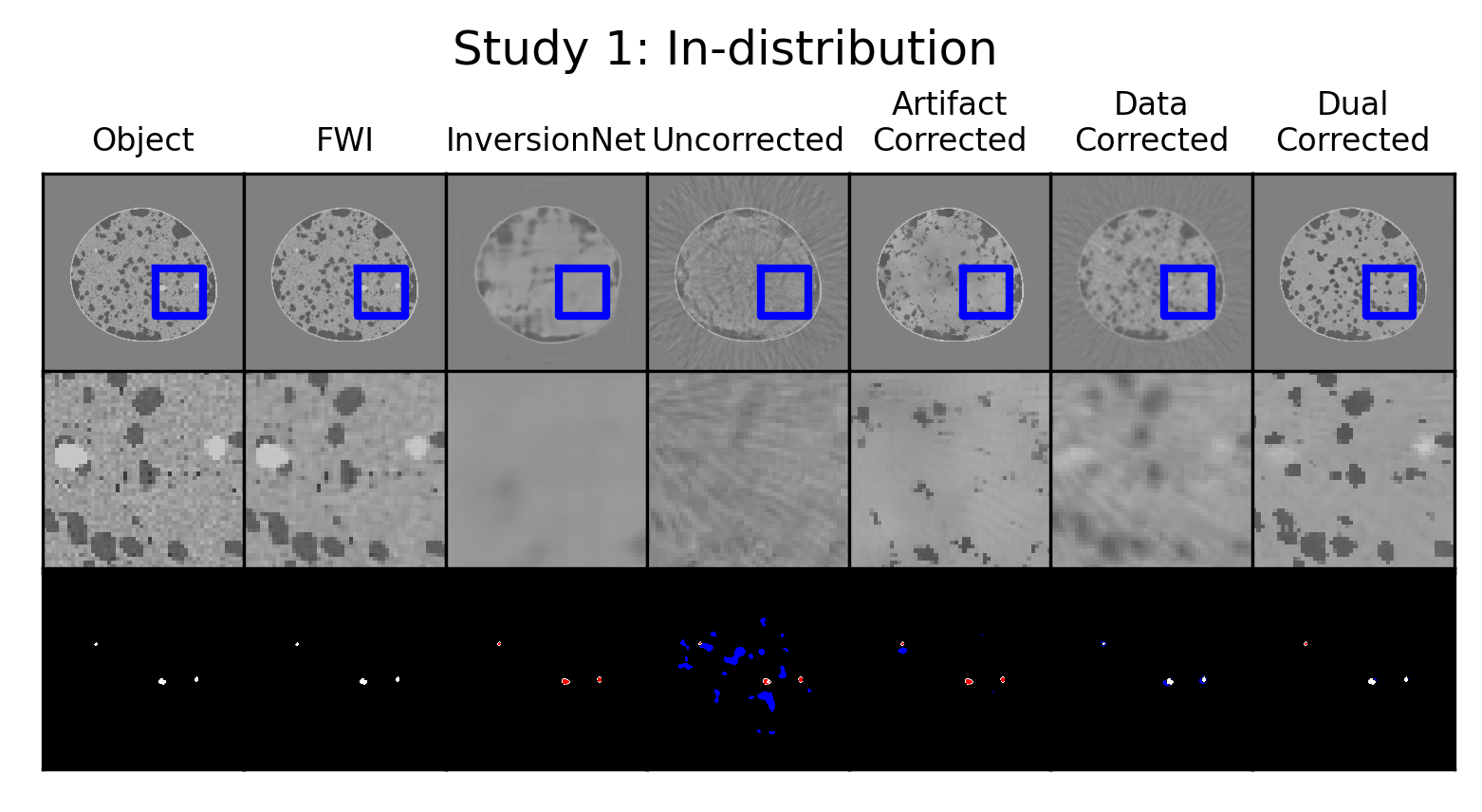}
        \caption{Study 1: In-distribution assessment. Example reconstruction of a Type D breast from the testing set. The top row represents the true object and its estimates reconstructed using each method. The middle row is a zoomed-in portion of the region outlined in the blue square. The grayscale range is [1400--1600] mm/$\mu$s. The bottom row displays the learned tumor segmentation for each reconstruction method. True positives (detections) are shown in white, true negatives are shown in black, false negatives are shown in red, and  false detections are shown in blue. The threshold for tumor detection was fixed for all reconstruction methods and chosen corresponding to the upper left corner of the ROC curve, which allows to maximize the balanced accuracy of the observer. The \emph{Data Corrected} method produced reconstructed images with the most accurate tumor segmentation out of those produced by all the learned methods. }
        \label{fig:example_in}
    \end{figure}

\subsection{Study 2: Out-of-distribution assessment}

\subsubsection{Qualitative assessment}
\begin{figure}
\centering
\includegraphics[width = \textwidth, trim = {0cm 0cm 0.5cm 0.7cm}, clip]{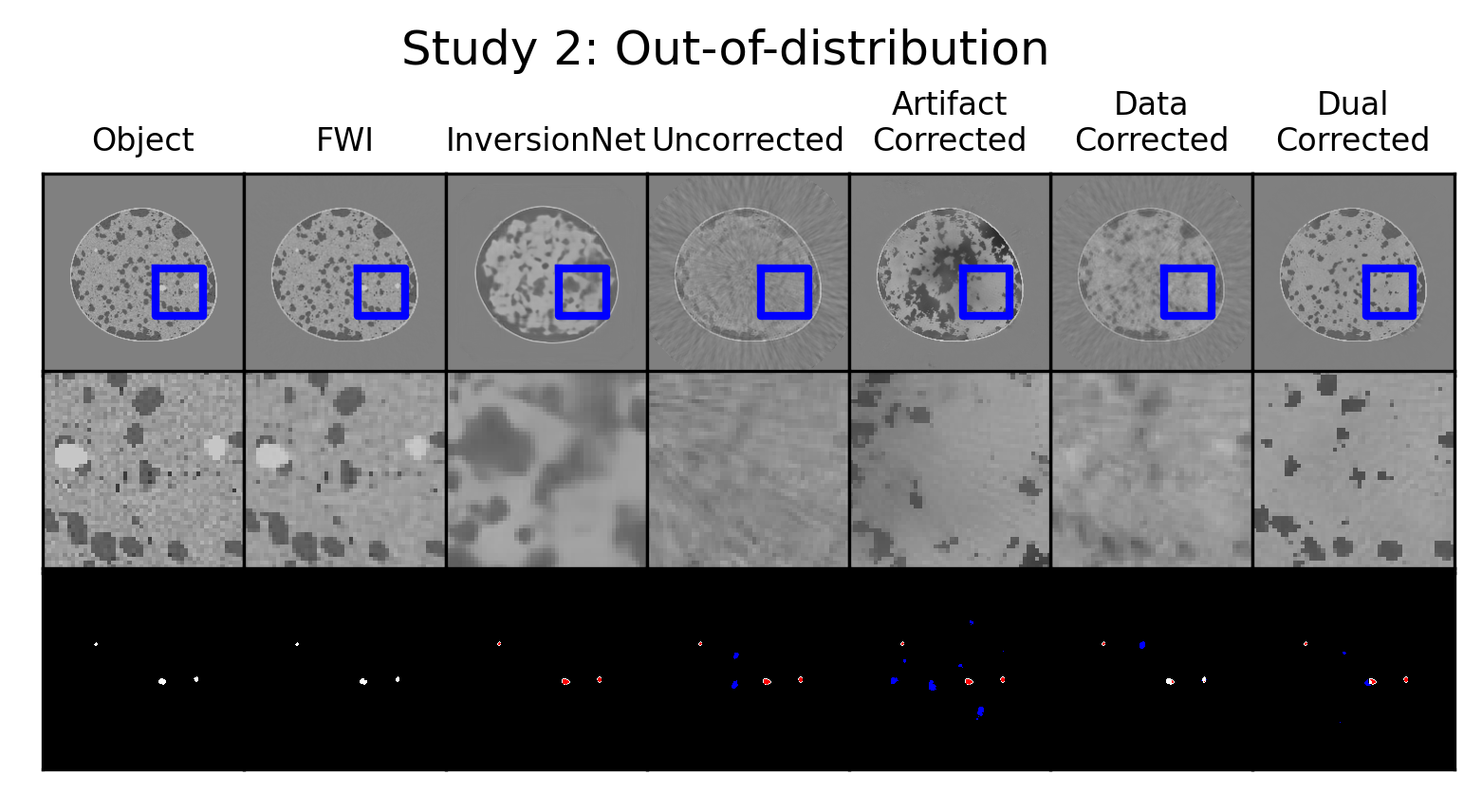}
\caption{Study 2: Out-of-distribution assessment. Example reconstruction of a Type D breast from the testing set.  The grayscale range is [1400--1600] mm/$\mu$s. While visually plausible, the \emph{InversionNet} reconstruction presents highly hallucinated internal structures and does not resemble the true object. The \emph{Artifact Correction} method produced a very realistic image; however, internal structures are also hallucinated and quantitative SOS values are biased towards a lower breast density, which is representative of the objects in the testing set.} 
\label{fig:example_out}
\end{figure}

Figure \ref{fig:example_out} features the same Type D example from the testing set that is also featured in Fig. \ref{fig:example_in}. Here, the \emph{FWI} and \emph{Uncorrected} reconstructions are the same in both figures. The \emph{InversionNet} reconstruction, while correctly capturing the shape of the breast, presents  hallucinated tissue structures typical of the lower density breast types used in the training set. The \emph{Artifact Correction} reconstruction does share some features in common with the true object, particularly those strongly informed from the starting \emph{Uncorrected} image. However, large portions of the \emph{Artifact Corrected} image in the interior of the breast have been altered and the resulting image seems to correspond to a lower density breast, such as the ones in the training set. The \emph{Data Corrected} reconstruction does not exhibit hallucinations or alterations to the image and remains consistent with results from Study 1, despite not being trained on Type D examples. Similarly, the \emph{Dual Corrected} reconstruction resembles a Type D breast and  seems largely consistent with results from Study 1. However, the \emph{Dual Corrected} reconstruction demonstrates reduced accuracy in the resulting tumor segmentation and the removal of key tumor features.

\subsubsection{Quantitative assessment}
\begin{figure}
\centering
\includegraphics[width = 0.62\textwidth, trim = {0cm 0cm 8cm 1.5cm}, clip]{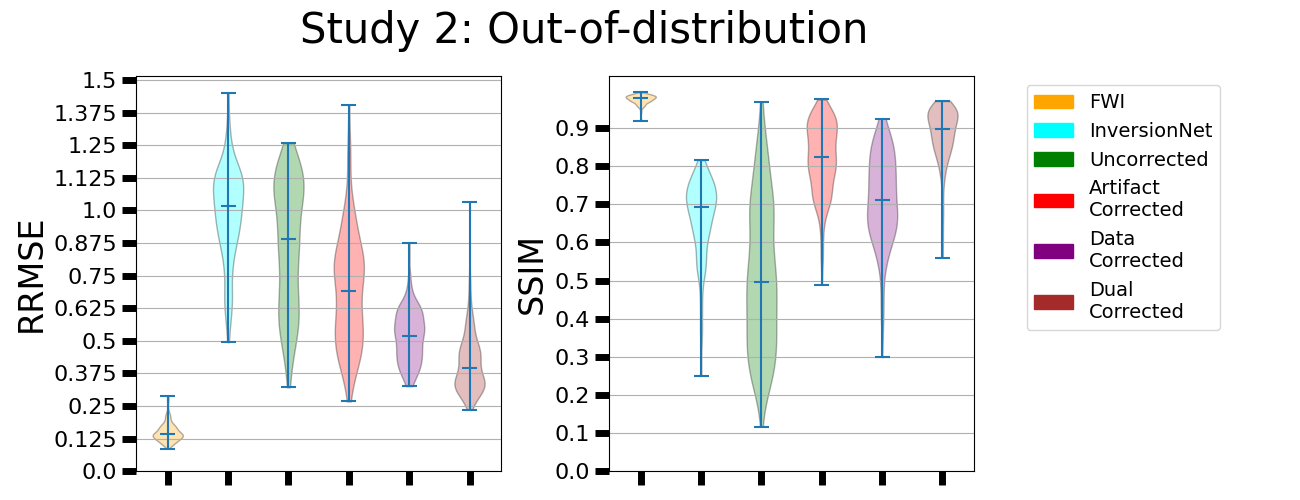}\includegraphics[width = 0.38\textwidth,  trim = {0cm 0cm 0cm 1.32cm}, clip]{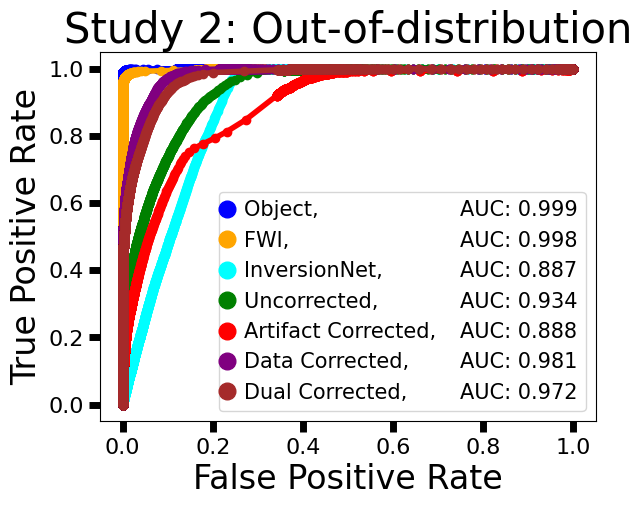}
\caption{Study 2: Out-of-distribution assessment. Violin plots for RRMSE (left), SSIM (center), and ROC curves (right) computed using reconstructed images from the testing set. All the learned reconstruction methods demonstrate a slight degradation in RRMSE compared to the results in Study 1 for in-distribution assessment. This increase in RRMSE is larger for  the \emph{InversionNet} and \emph{Artifact Corrected} methods. Notably, there isn't a large decrease in SSIM between Study 1 and Study 2. The \emph{Artifact Corrected} demonstrates a significant drop in observer performance compared to the other methods in this study. }
\label{fig:violin_plots_out}
\end{figure}

The ensemble RRMSE and SSIM violin plots and ROC curves computed using reconstructed images from the testing set are displayed in Fig \ref{fig:violin_plots_out}. 

Note that every reconstruction method, except the \emph{Uncorrected} method, demonstrates a statistically significant increase in RRMSE (p values $< 10^{-5})$. The reason why the\emph{Uncorrected}  method achieves better performance in terms RRMSE compared to Study 1 (p-value $\approx 10^{-12}$) is because, despite the higher spatial heterogeneity Type D breast, the average SOS value is closer to that of water compared to less dense breasts those average SOS value is significantly smaller than that of water. Note that the \emph{InversionNet} method demonstrates a particularly large increase in RRMSE and even performs worse than the \emph{Uncorrected} method. 
Additionally, all methods, except for the \emph{Artifact Corrected} method, produced SOS estimates with a statistically significant increase in SSIM (p-values $< 0.003$), while the \emph{Artifact Corrected} method demonstrates an increase in SSIM that is not statistically significant (p-value $= 0.107$). These statistically significant changes can largely be explained by the testing sets in Studies 1 and 2 being based on different distributions of images. In particular, Type D breasts are often smaller than other breast types, have a higher average SOS value (glandular tissue has faster SOS than fat), and have stronger acoustic heterogeneity.

The ROC analysis in Fig. \ref{fig:violin_plots_out} displays that the \emph{Artifact Corrected} and \emph{Dual Corrected} methods produce SOS estimates of lower task-based quality compared to the \emph{Uncorrected} and \emph{Data Corrected} approaches, respectively. In particular, the ROC curve for the \emph{Artifact Corrected} approach exhibits a very prominent cusp in its upper left corner not present in the other ROC curves. This suggests that applying artifact correction to out-of-distribution data can corrupt task-relevant information and hinder task performance.

\subsubsection{Discussion}
This study serves to assess the generalization power of these reconstruction methods applied to out-of-distribution data.  Particularly, this study demonstrates that purely data-driven and artifact correction methods are prone to hallucinating and altering features in the reconstructed images, possibly introducing bias from the training set. Whereas an approach utilizing correction in the measurement seems to avoid this bias and is more robust with respect to domain shift. 

\subsection{Study 3: Effect of training data paucity}

\subsubsection{Qualitative assessment}
\begin{figure}
\centering
\includegraphics[width = \textwidth, trim = {0cm 0.5cm 0cm 0.7cm}, clip]{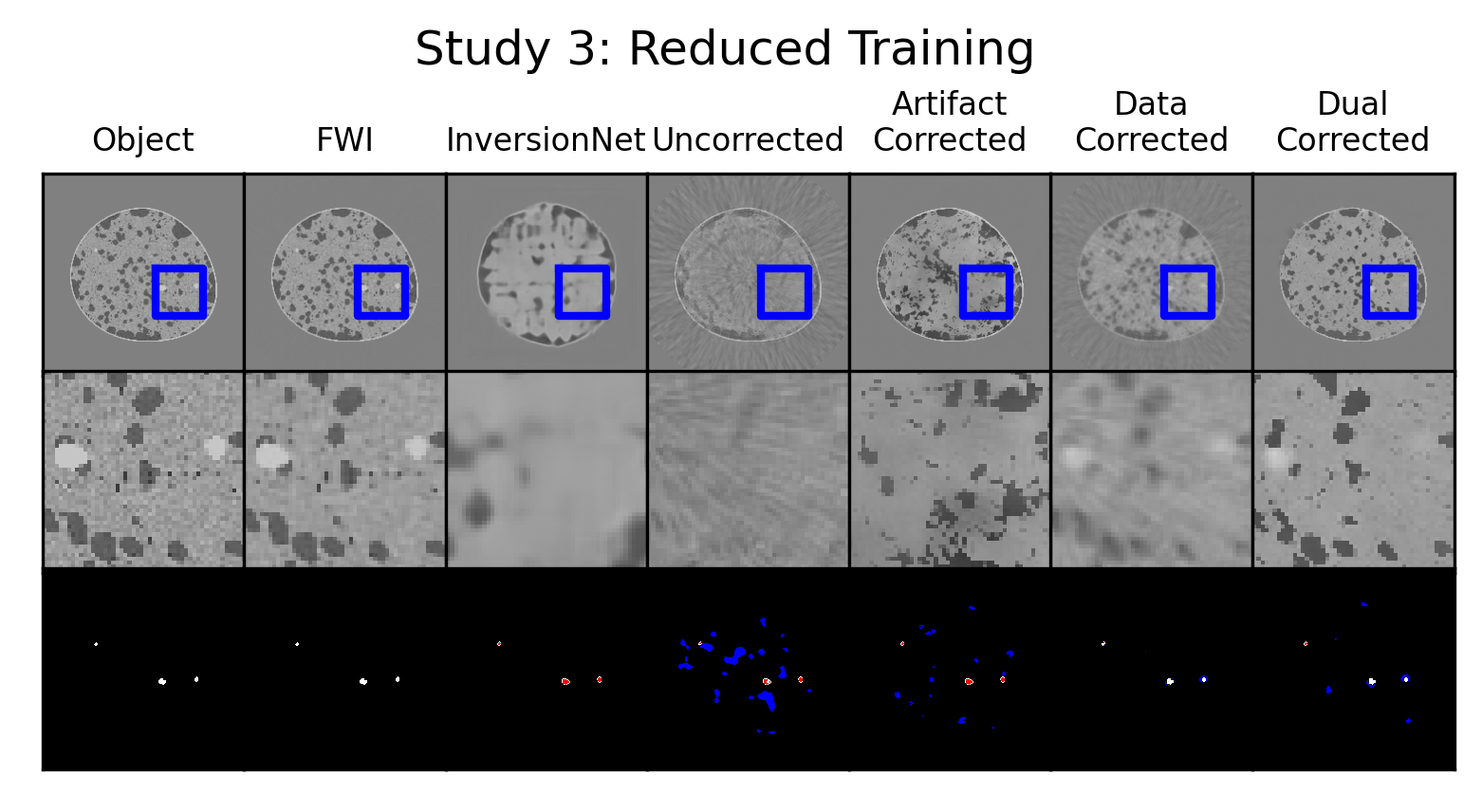}
\caption{Study 3: Reduced training set size. Example reconstruction of a Type D breast from the testing set. The grayscale range is [1400--1600] mm/$\mu$s.  The \emph{Artifact Corrected} and \emph{InversionNet}reconstructions have very apparent hallucinations compared to Study 1. On the contrary, the \emph{Data Corrected} reconstruction has comparable perceptual quality to the reconstruction shown in Study 1.}
\label{fig:example_red}
\end{figure}

The example reconstruction in Fig. \ref{fig:example_red} illustrates the effects of a reduced training set for each reconstruction method. The \emph{InversionNet} reconstructions do not seem to represent a realistic breast compared to the reconstructions in Studies 1 and 2. The \emph{Artifact Corrected} reconstructions seem perceptually realistic but the interior structures do not correspond to those of the true object. The \emph{Data Corrected} and \emph{Dual Corrected} reconstructions present the same visual quality as those in Study 1 and lead to accurate tumor detection, indicating that such approaches can perform well also in training data-scarce applications. 

\subsubsection{Quantitative assessment}
\begin{figure}
\centering
\includegraphics[width = 0.62\textwidth, trim = {0cm 0cm 8cm 1.5cm}, clip]{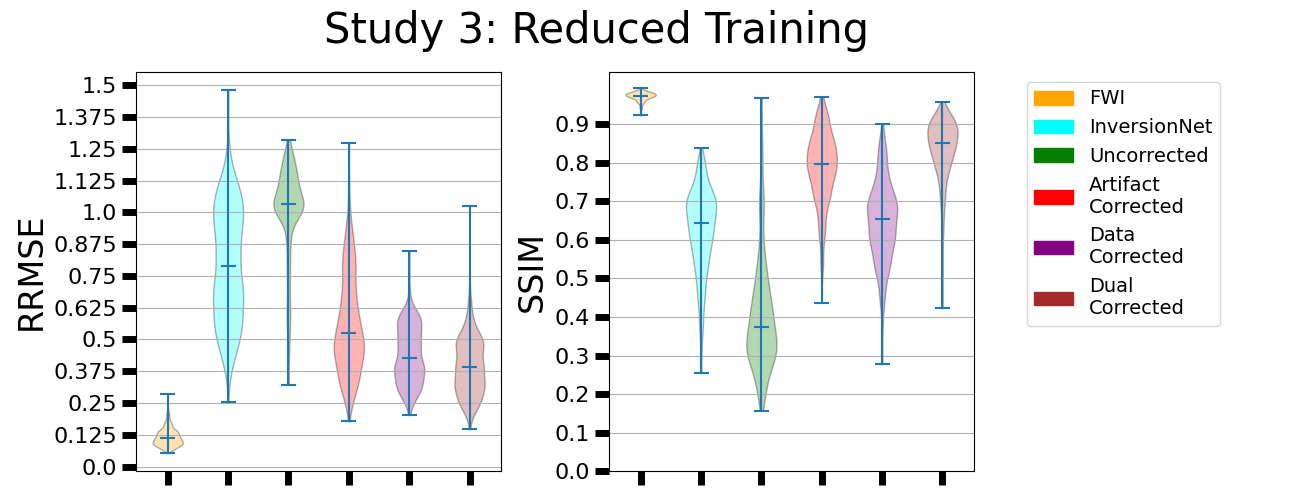}\includegraphics[width = 0.38\textwidth,  trim = {0cm 0cm 0cm 1.32cm}, clip]{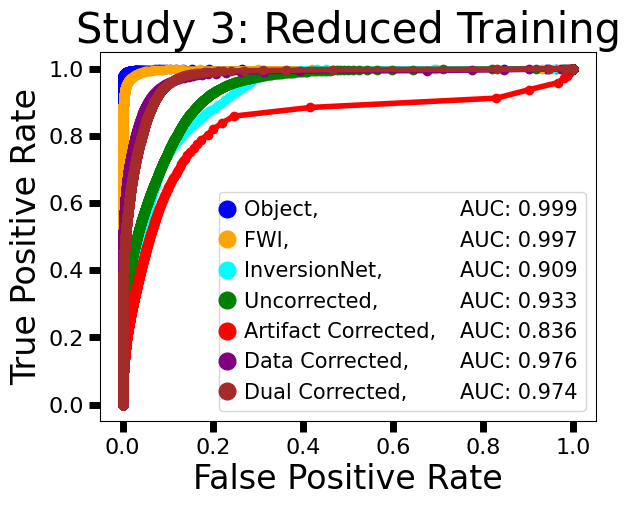}
\caption{Study 3: Reduced Training. Violin plots for RRMSE (left), SSIM (center), and ROC curves computed on reconstructed images from the testing set. The \emph{Artifact Corrected} method demonstrates an increase in RRMSE compared to the results in Study 1. The learned reconstruction methods incorporating the Born approximation demonstrate a slight decrease in SSIM compared to the results in Study 1. The \emph{InversionNet} approach demonstrates a slight decrease in RRMSE and an increase SSIM compared to the other approaches with very poor task performance as demonstrated by AUC.}
\label{fig:violin_plots_red}
\end{figure}

The ensemble RRMSE and SSIM violin plots and ROC curves computed using reconstructed images from the testing set are displayed in Fig. \ref{fig:violin_plots_red}.  While the testing set used in this study is larger than that used in Study 1, they are both sampled from the same distribution. Therefore, the intrinsic difficulty of the test set used here is the same as the one of Study 1, as demonstrated by the fact that, for the \emph{FWI} and \emph{Uncorrected} reconstruction methods, there is no statistical difference between RRMSE or SSIM reported here and those reported in Study 1 (all p-values > 0.48).
The \emph{InversionNet} method demonstrates a significant increase in RRMSE (p-value $<10^{-4}$) and a slight decrease in SSIM (p-value = 0.0285).  The \emph{Artifact Corrected} method demonstrates a significant increase in RRMSE (p-value $< 10^{-5}$) and a decrease in SSIM (p-value = 0.007). The \emph{Data Corrected} method does not demonstrate a significant change in RRMSE (p-value $= 0.923$)  or SSIM (p-value $= 0.929$). The \emph{Dual Corrected} method demonstrates a significant increase in RRMSE (p-value = $0.0009$) and a slight decrease in SSIM corresponding to a p-value of 0.0207.

The ROC analysis in Fig. \ref{fig:violin_plots_red} shows that the task performance of the \emph{Artifact Corrected} method is particularly degraded as the size of the training set is reduced, while the other methods achieve very similar task performance compared to Study 1.

\subsubsection{Discussion}

This study serves to assess the effects of a reduced training set on the performance of each reconstruction method employing machine learning. The purely data-driven reconstruction method suffers a large reduction in accuracy and visual appearance when trained using fewer examples.  This indicates that the accuracy of data-driven approaches is heavily dependent on the amount of available training data and is not well suited for situations with limited training data. Similarly, the approaches utilizing artifact correction face a reduction in accuracy, although less severe than the pure data-driven approach. On the contrary, the approach only utilizing data correction demonstrates comparable performance between Study 1 and Study 3. The robustness of the data correction approach with respect to the availability of limited training examples can be attributed to two distinct reasons. On the one hand, USCT measurement data (pressure traces) is much larger and richer compared to SOS images. Therefore, even though the number of training examples is the same, the loss function in the USCT measurement domain is more informative and better constraints the weight of the \emph{Data Correction} network compared to the image domain loss used to train the \emph{Artifact Corrected} network. On the other hand, performing the Born inversion after applying the learned method to the measurement data can effectively filter any inconsistency between the output of the network and the range of the imaging operator. 

\subsection{Study 4: Robustness with respect to measurement noise}

\subsubsection{Qualitative assessment}
\begin{figure}
\centering
\includegraphics[width = \textwidth, trim = {0cm 0.5cm 0cm 0.7cm}, clip]{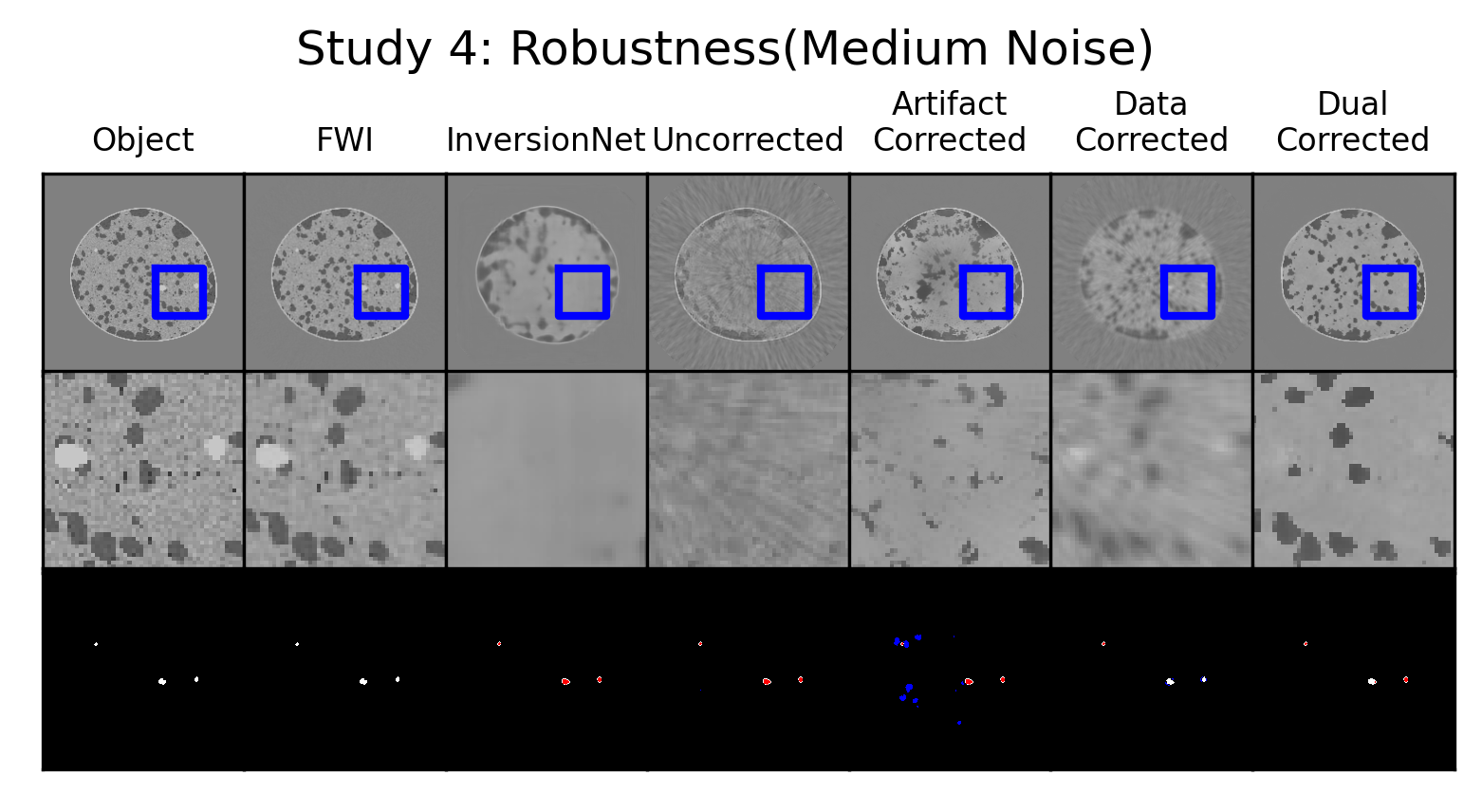}
\includegraphics[width = \textwidth, trim = {0cm 0.5cm 0cm 1.5cm}, clip]{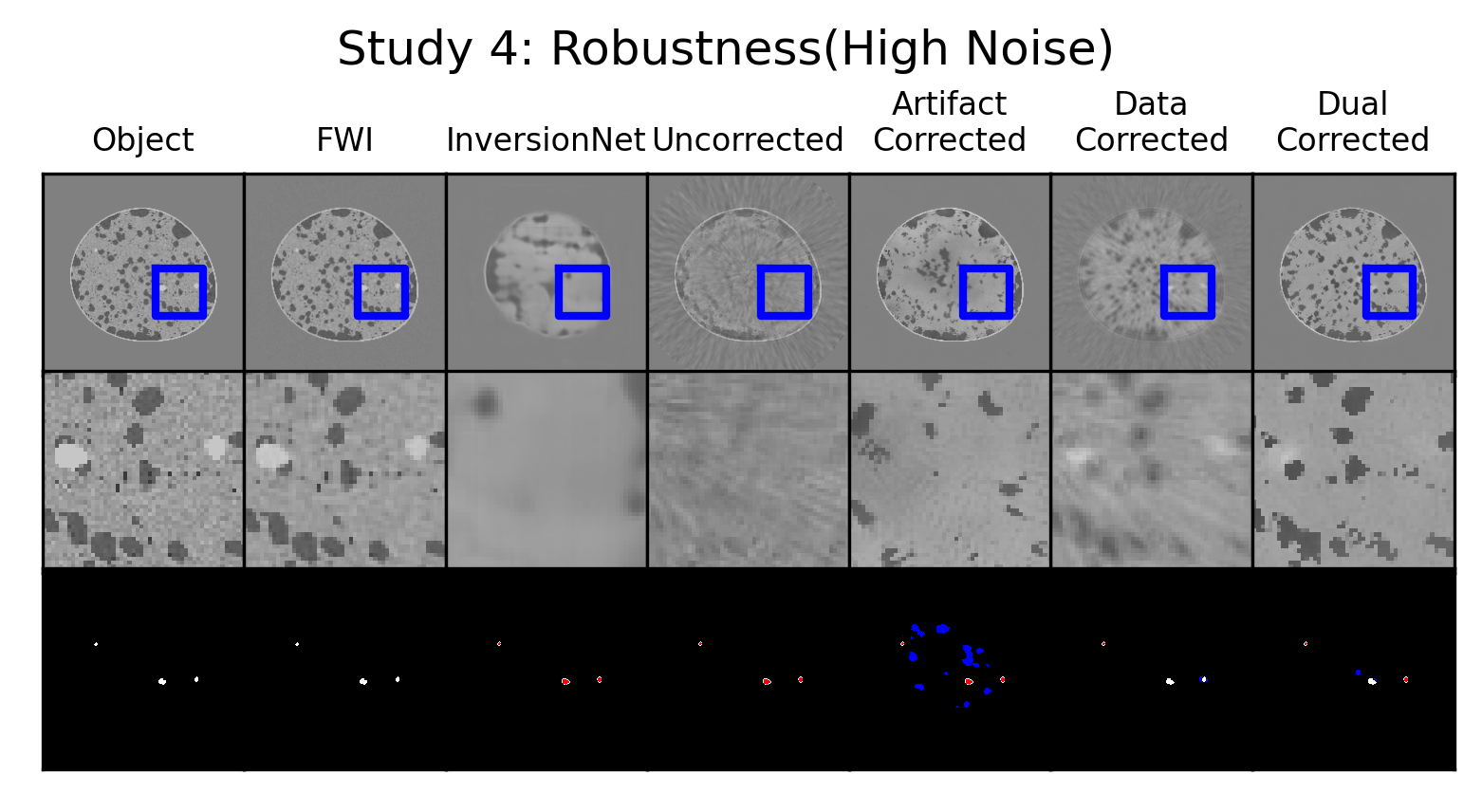}     
\caption{Study 4: Robustness (top panel: Medium Noise, bottom panel: High Noise). Example reconstruction of a Type D breast from the testing set. The grayscale range is [1400--1600] mm/$\mu$s. The \emph{Data Corrected} reconstruction displays greater artifacts due to model mismatch and blurring with increased noise while the \emph{Artifact Corrected} and \emph{Dual Corrected} methods remain largely stable.}
\label{fig:example_rob}
\end{figure}

Figure \ref{fig:example_rob} displays an example of a reconstructed image (a type D NBP from the testing set) in the medium (top) and high (bottom) measurement noise regimes, with the training and testing noise matched. In both noise regimes, the \emph{InversionNet} reconstruction presents hallucinated tissue structures, and its visual appearance is significantly different from the image shown in Study 1. The \emph{Artifact Corrected} reconstructions display perceptual similarity to the true objects, but suffer from few noticeable hallucinations that were not present in the image shown in Study 1.  The artifacts due to model mismatch in the \emph{Data Corrected} reconstructions are larger compared to Study 1, and more severe in the high noise regime, with blurring around the boundary of the breast.  While the accuracy of the \emph{Data Corrected} degrades as the noise level increases, tumor regions are still accurately detected and segmented. These larger artifacts from the \emph{Data Corrected} method further impact the \emph{Dual Corrected} approach and lead to minor changes in local features, for example, the changes in shape boundaries featured in the zoomed-in portion of Fig. \ref{fig:example_rob}.

\subsubsection{Quantitative assessment}

\begin{figure}
    \centering
    \includegraphics[width=\textwidth, trim = {0cm 0.9cm 0cm 0.0cm}, clip]{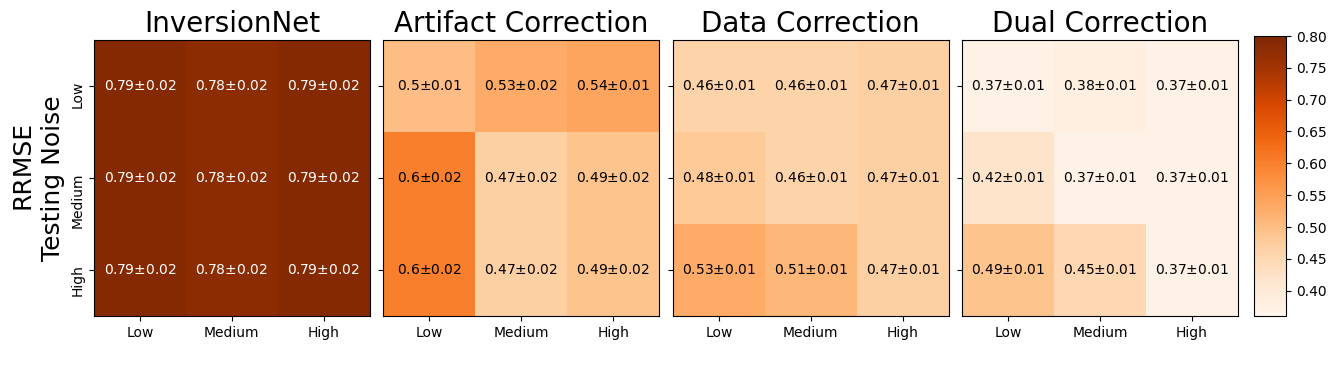}\\
    \includegraphics[width=\textwidth,  trim = {0cm 0.9cm 0cm 0.82cm}, clip]{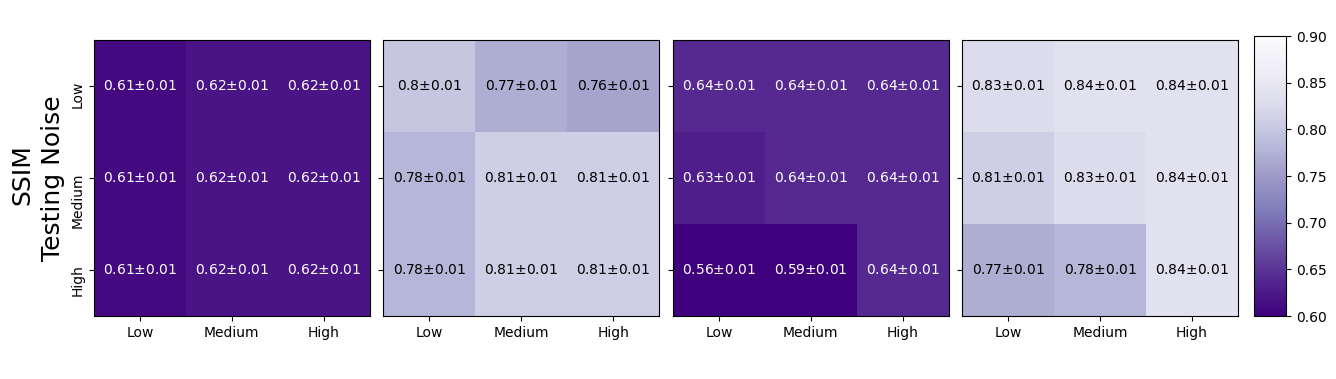}\\
    \includegraphics[width=\textwidth, trim = {0cm 0.0cm 0cm 0.82cm}, clip]{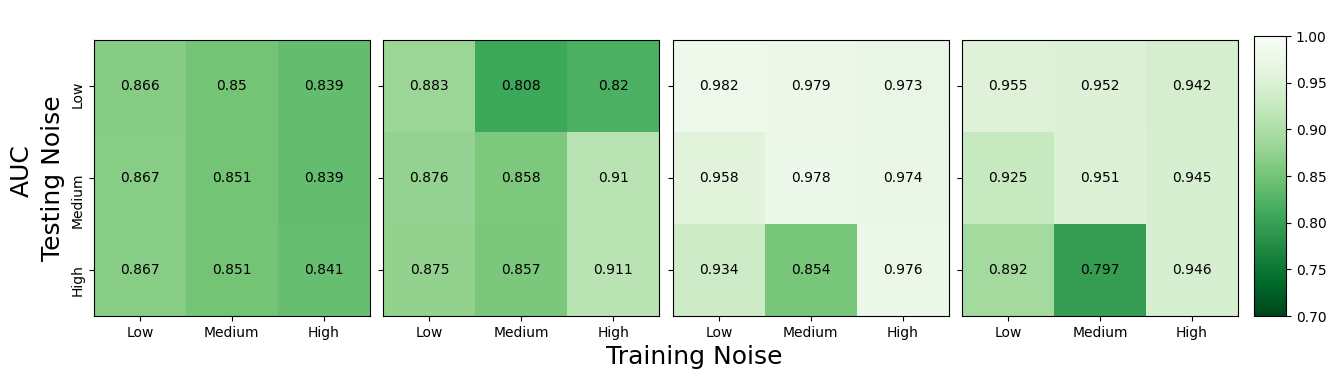}
    
    \caption{Study 4: Robustness (top row: RRMSE confusion plots for each method, middle row: SSIM confusion plots, bottom row: ROC AUC confusion plots). The RRMSE and SSIM confusion plots display the mean values from the testing set and additionally display the 95 percent confidence interval for the estimated mean. The correction methods demonstrate a greater level of robustness across noise levels when trained with a higher noise level. }
    \label{fig:confusion_plots}
\end{figure}

The confusion plots for RRMSE, SSIM, and AUC for the ROC curve from each method when trained and tested at each of the three noise levels are shown in Fig. \ref{fig:confusion_plots}. The \emph{InversionNet} method demonstrates poor performance regardless of training and testing noise across all three metrics. The \emph{Artifact Correction} demonstrates the highest accuracy, in terms of all three metrics, when trained and tested on the same level of noise and a decrease in performance when tested on a different level of noise. The \emph{Data} and \emph{Dual Correction} methods demonstrate high levels of robustness and improved performance when trained on a higher level of noise and tested on lower levels of noise. Similarly, the \emph{Data} and \emph{Dual Correction} methods demonstrated reduced accuracy for all three metrics when trained on a low level of noise and tested on a high level of noise. Note that the task based assessment of image quality follows the same general trends as the image quality metrics with changes in level of measurement noise.

\subsubsection{Discussion}

This study assesses the robustness of each reconstruction method to increased measurement noise. In particular, this study demonstrates a greater level of robustness in data correction methods trained on a higher level of noise. Furthermore, the purely data-driven and artifact correction approaches do not demonstrate this same level of robustness between training and testing noise levels, but do demonstrate a substantial reduction in accuracy as the matched level of training and testing noise increased.  Combining both forms of correction still leads to the most accurate reconstructions in terms of SSIM and RRMSE  while data correction alone leads to the best task performance as quantified by AUC.

\section{Conclusion}\label{sec:conclusion}

This work systematically compares multiple hybrid image reconstruction methods combining an approximated forward model and machine learning with applications to ultrasound computed tomography (USCT).  These hybrid reconstruction  approaches include the application of neural networks for data correction in the measurement domain and artifact correction in the image domain. Full-waveform inversion, a model-based image reconstruction using the accurate forward model, serves as a reference and approaches using only approximate physics (Born inversion) and purely data-driven methods are also considered. To compare the accuracy and performance of these approaches, four virtual imaging studies were performed. The first study provides an assessment for in-distribution measurement data, that is when the training and testing sets are sampled from the same distribution of numerical breast phantoms representative of all breast density types. The second study assessed the methods on out-of-distribution measurement data, where the training set consisted of low to moderate breast density types while the testing set consisted of extremely dense breasts. The third study assessed the performance of the methods when training data is limited. The fourth study assessed each reconstruction method's robustness to measurement noise. 

These studies demonstrate that machine learning methods can effectively mitigate modeling errors induced by the use of a simplified, linearized model of wave propagation. The learned correction approaches not only improved the accuracy of USCT reconstruction measured in terms of root mean square error (RRMSE) and structural similarity (SSIM), but also increased task performance (area under the curve) in a tumor detection task using numerical observers. In particular, all the learned correction approaches greatly outperformed a purely-data-driven method, that did not incorporate physics, under all three metrics.  Among the learned correction approaches,  data correction in the measurement domain demonstrated minor visual artifacts but the highest task performance, best generalizability to out-of-distribution data, and a high level of robustness across testing noise levels when trained on a high level of noise. Artifact correction in the image domain approaches was susceptible to a heavy bias towards training data, the presence of hallucinations in reconstructed images and the loss of task-relevant features; however, they also demonstrated a greater robustness towards increased levels of matched training and testing noise, but were not robust to mismatched levels of noise. Combining both forms of correction, in a dual corrected method, resulted in the most accurate reconstructed images, in terms of SSIM and RRMSE, at the cost of a slight reduction in task performance as compared to the data corrected method and inherited the robustness across testing noise levels when trained a high level of noise from the data correction method.

In summary, these studies indicated that incorporating physics, even a simplified or physical model, can lead to improved accuracy, stability to out-of-distribution data, and task performance compared to purely data-driven approaches. However, the accuracy of these corrected approaches is still severely reduced compared to approaches incorporating accurate but computationally expensive physics models. This suggests expensive traditional model-based approaches are preferable when high accuracy is a much greater priority compared to speed in reconstruction and reconstructions utilizing learned correction may be instead useful for providing an initial guess for these model-based approaches.

Future work may investigate the use of different simplified physics models or physical models with underlying uncertainties and alternative reconstruction methods, such as filtered back projection. Additionally, as a first step towards a clinical application to ring-array breast USCT systems, the methods considered here may be further extended to incorporate simultaneously learned corrections for both unmodeled three-dimensional wave propagation effects \cite{LiVillaAnastasio24p}and the use of the Born approximation. 
Virtual imaging studies employing a high-fidelity model of the data acquisition process \cite{LiVillaDuricEtAl22,LiVillaDuricEtAl23}, which accounts for three-dimensional wave propagation physics and transducer focusing properties, may then provide a systematic assessment of such learned correction in clinically relevant scenarios.

\section*{Disclosures}
The authors declare that they have no conflict of interest.

\section*{Code and Data Availability}
The code and the training data used in this study are available upon request, please contact Umberto Villa at \linkable{uvilla@oden.utexas.edu}.

\section*{Acknowledgments}
This work was funded in part by the National Institutes of Health under awards R01 EB034261 \& R01EB028652, and in part by Los Alamos National Laboratory (LANL) through the Center for Space and Earth Science and Laboratory Directed Research and Development program under Grant 20200061DR. 

\bibliographystyle{spiejour}
\bibliography{local} 
\end{document}